\documentclass{aa}  

\usepackage{subcaption}
\usepackage{xcolor}
\usepackage{graphicx}
\usepackage{txfonts}
\usepackage{pdflscape}
\usepackage{natbib}
\usepackage[colorlinks=true,citecolor=blue]{hyperref}
\bibpunct{(}{)}{;}{a}{}{,}
\newcommand{\orcid}[1]{\href{https://orcid.org/#1}{\includegraphics[width=10pt]{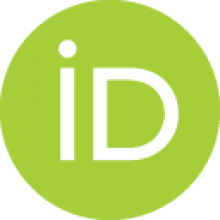}}}

\usepackage{placeins}

\defcitealias{K18}{K18}

\begin{document} 

\title{Late-time evolution of the interacting stripped-envelope supernova 2017dio}

\author{C. Humina
          \inst{1}\orcid{0009-0008-2621-9146}
          \and
          C.~P.~Gutiérrez
          \inst{2,3}\orcid{0000-0003-2375-2064}
          \and
          H. Kuncarayakti
          \inst{1,6}
          \and
          T. Nagao
          \inst{1,4}
          \and
          S. Gonz\'alez-Gait\'an
          \inst{5}\orcid{0000-0001-9541-0317}
          \and
          T. Kangas
          \inst{6,1}\orcid{0000-0002-5477-0217}
          \and
          T. Kravtsov
          \inst{6,1}\orcid{0000-0003-0955-9102}
          \and
          K. Maeda
          \inst{7}\orcid{0000-0003-2611-7269}
          \and
          S. Mattila
          \inst{1,8}\orcid{0000-0001-7497-2994}
          \and
          T. M. Reynolds
          \inst{1,9,10}\orcid{0000-0002-1022-6463}
          \and
          L. Tartaglia
          \inst{11}\orcid{0000-0003-3433-1492}
          \and
          S. Benetti
          \inst{12}\orcid{0000-0002-3256-0016}
          \and
          M. Fraser
          \inst{13}
          \and
          C. Gall
          \inst{14}\orcid{0000-0002-8526-3963}
          \and
          P. Lundqvist
          \inst{15}\orcid{0000-0002-3664-8082}
          \and
          A. Morales-Garoffolo
          \inst{16}\orcid{0000-0001-8830-7063}
          \and
          M. Stritzinger
          \inst{17}\orcid{0000-0002-5571-1833}
          }

   \institute{
   Department of Physics and Astronomy, University of Turku, FI-20014, Finland\\
   \email{jahumi@utu.fi}
   \and
   Institute of Space Sciences (ICE, CSIC), Campus UAB, Carrer de Can Magrans, s/n, E-08193 Barcelona, Spain \\
   \email{cgutierrez@ice.csic.es}
   \and
   Institut d'Estudis Espacials de Catalunya (IEEC), Edifici RDIT, Campus UPC, 08860 Castelldefels (Barcelona), Spain
   \and 
   National Astronomical Observatory of Japan, National Institutes of Natural Sciences, 2-21-1 Osawa, Mitaka, Tokyo 181-8588, Japan
   \and
   Instituto de Astrof\'isica e Ci\^encias do Espaço, Faculdade de Ci\^encias, Universidade de Lisboa, Ed. C8, Campo Grande, 1749-016 Lisbon, Portugal
   \and 
   Finnish Centre for Astronomy with ESO (FINCA), FI-20014 University of Turku, Finland
   \and
   Department of Astronomy, Kyoto University, Kitashirakawa-Oiwake-cho, Sakyo-ku, Kyoto 606-8502, Japan 
   \and
   School of Sciences, European University Cyprus, Diogenes Street, Engomi, 1516 Nicosia, Cyprus
   \and
    Cosmic Dawn Center (DAWN)
   \and
   Niels Bohr Institute, University of Copenhagen, Jagtvej 128, 2200 København N, Denmark
   \and
   INAF -- Osservatorio Astronomico d’Abruzzo, via Mentore Maggini snc, I-64100, Teramo, Italy
   \and
   NAF-Osservatorio Astronomico di Padova, vicolo dell’Osservatorio 5, 35122 Padova, Italy
   \and
    DARK, Niels Bohr Institute, University of Copenhagen, Jagtvej 155A, DK-2200 Copenhagen N, Denmark
   \and
   School of Physics, University College Dublin, LMI Main Building, Beech Hill Road, Dublin 4, D04 P7W1, Ireland
   \and
    The Oskar Klein Centre, Department of Astronomy, Stockholm University, AlbaNova SE-10691, Stockholm, Sweden
   \and
   Department of Applied Physics, School of Engineering, University of Cádiz, Campus of Puerto Real, E-11519 Cádiz, Spain
   \and
   Department of Physics and Astronomy, Aarhus University, Ny Munkegade 120, DK-8000 Aarhus C, Denmark\label{uni:aarhus}
   }

   \date{}

\abstract{  
{The discovery of stripped-envelope core-collapse supernovae (SNe) interacting with dense circumstellar medium (CSM) challenges our current understanding of massive star evolution.}
{We present late-time observations of the interacting Type Ic SN~2017dio and investigate its mass-loss mechanism and progenitor channel.}
{We analysed late-time spectra and light curves that are dominated by a strong ejecta–CSM interaction and reveal a slow photometric evolution primarily powered by this interaction. We examined the CSM and the source of the infrared (IR) excess by modelling the radiation produced by the ejecta–CSM interaction and the IR echo from circumstellar dust. In addition, we studied the temporal evolution of spectral features, with a particular emphasis on the H$\alpha$ emission line.}
{From the combined analysis of the light curves and spectral properties, we infer that the peak mass-loss rate for the CSM reaches $\sim0.2$ $\mathrm{M_{\odot}~yr^{-1}}$ and  that the typical value over most epochs is $\sim0.06$ $\mathrm{M_{\odot}~yr^{-1}}$. The nearby CSM was formed over a period of 4 to 65 years before the explosion. The CSM radius begins at $\sim 1.3\times10^{15}$ cm. The IR excess identified in the light curves is consistent with the radiation from dust with a mass increasing from $\sim0.001$ to $\sim0.005~ \mathrm{M_{\odot}}$ in the case of carbon dust or $\sim0.005$ to $\sim0.02~ \mathrm{M_{\odot}}$ in the case of silicate dust.}
{From IR echo modelling, we estimate an upper limit on the dust mass of $4\times10^{-5}~\mathrm{M_{\odot}}$, which implies an SN progenitor mass-loss rate of $2.4\times10^{-5}~\mathrm{M_{\odot}~yr^{-1}}$ at the dust evaporation radius determined by the SN peak luminosity (0.017 pc for carbon dust, corresponding to mass loss $\sim170$ years before the explosion). This implies a very rapid increase in the mass-loss rate ahead of the explosion.

Although the progenitor of SN~2017dio has lost its helium envelope, it interacted with a hydrogen-rich CSM formed shortly before the explosion, strongly suggesting that this material originated from a companion star rather than the progenitor itself.}

{}
}

   \maketitle
%

\section{Introduction}
Core-collapse supernovae (SNe) signal the death of massive stars. SNe are classified according to the presence or absence of certain elements in their spectra (and, in some cases, by their light-curve shapes, e.g. Types IIP and IIL; \citealt{Barbon79}). Pre-explosion mass loss plays a crucial role in shaping the resulting SN type, and SNe that have lost most or all of their hydrogen and helium envelopes are called stripped-envelope SNe (SESNe). This includes Types IIb (whose progenitors have lost most of their hydrogen envelope), Ib (whose progenitors have lost all of their hydrogen envelope), Ic (whose progenitors have lost both their hydrogen and helium envelopes), and Ic-BL, whose lines are similar to those of Type Ic SNe but broader \citep[see e.g.][]{Filippenko, Modjaz19}. Some core-collapse SNe have strong interactions with the surrounding circumstellar medium (CSM). The CSM is thought to originate from the mass loss of the progenitor system in the final moments before the SN explosion. Interaction with CSM can create features observable in both spectra and light curves, and it can sometimes even hide the underlying SN type. Type IIn (`n' stands for narrow lines) are SNe that show narrow Balmer emission lines in the spectra, which are interpreted to be a product of the SNe interacting with H-rich CSM \citep{IIn}.

Some SESNe have been found to be interacting with a significant amount of CSM as well. Type Ibn SNe interact with He-rich CSM (\citealt{Pastorello}, \citealt{Foley_2007}) and Icn with O/C-rich CSM \citep{Icn}. A classification of Type Ien has also been proposed for SN~2021yfj, which showed narrow emission lines in Si, S, and Ar (\citealt{Ien}, \citealt{Schulze25}). Another rare class of interacting SESNe is composed of events that start to interact with H-rich CSM at a certain point of their evolution and evolve into Type IIn SNe. This includes SN~2014C \citep{2014c}, 2017ens \citep{2017ens}, SN~2017dio (\citealt{K18}, hereafter \citetalias{K18}), SN~2018ijp \citep{Tartaglia}, and SN~2019yvr \citep{yvr}.

The mass-loss mechanisms for SESNe are still uncertain. Current models argue that stripping can be done by stellar winds \citep{Groh2013} and by binary interaction \citep[see e.g.][]{bin,yoon2017}. Theoretical studies and direct observations of SESN progenitors both strongly indicate that the removal of hydrogen results from mass transfer in binary systems \citep[see e.g.][]{Ouchi,bin,2008ax}, but for helium-stripping, additional mass-loss mechanisms are needed \citep{Fang}. Possible candidates for helium-stripping are strong stellar winds from Wolf-Rayet-like stars \citep{Crowther2007}, although non-steady mass ejections have also been suggested as a way to strip helium (e.g. \citealt{Pyykkinen}).

SN~2017dio was the first Type Ic SN observed to show signs of interaction with H-rich CSM in early spectra. It was first classified as a Type Ic SN \citep{class_report}, but \citetalias{K18} found that SN~2017dio was a Type Ic interacting with H-rich CSM and that the interaction dominated the spectra later on. Due to the nature of the SN type classification, questions arise about what kind of progenitor channel is needed to cause the progenitor to lose most of its He envelope while still retaining the H-rich CSM close by. Additionally, \cite{Thevenot} found an infrared (IR) excess in SN~2017dio, suggesting a large amount of either newly formed or pre-existing dust. In this paper, pre-existing dust refers to dust formed in the material ejected by the SN progenitor prior to the explosion, i.e. in the CSM. 

Dust emission has been observed in many interacting SNe, particularly in Type IIn such as SN~2005ip (\citealt{Fox_2005ip}, \citealt{2005ip_1}, \citealt{2005ip}), SN~2006jd \citep{2005ip_1}, SN~2010jl \citep{2010jl_IR}, and SN~2017hcc (\citealt{2017hcc}, \citealt{2017hcc2}), as well as in Type Ibn and Icn SNe such as SN~2006jc \citep{2006jc} and SN~2023xgo \citep{2023xgo}. Signatures of dust have been observed in transitional objects, including SN~2014C \citep{2014c_Tinyanont} and SN~2017ens \citep{2017ens_IR}. The first detection of dust emission associated with a Type Ic SN was reported for SN~2020oi \citep{Rho_2021}, followed more recently by SN~2022jli \citep{2022jli}. If the IR excess arises from pre-existing dust, determining its properties would allow us to infer the properties of the associated CSM and, therefore, constrain the earlier mass-loss history of the progenitors of the SNe. In most cases, it is difficult to disentangle whether the dust is pre-existing or newly formed \citep{Gall2018}.

We studied the detailed properties of the CSM and the source of the IR excess by modelling the optical and IR light curves, as well as by studying the spectral features arising from CSM interaction. Throughout this work, we assume a flat $\Lambda$ cold dark matter universe with a Hubble constant of $H_0=70$\,km\,s$^{-1}$\,Mpc$^{-1}$ and $\Omega_\mathrm{m}=0.3$.
The paper is organised as follows. Section~\ref{sec:Obs_red} presents the observations and data reduction. Section \ref{host_gal} presents properties of the host galaxy. Sections \ref{sec:photometric_properties} and \ref{sec:spectroscopic_properties} present the photometric and spectroscopic properties, respectively. In Sects.~\ref{LC_modelling} and \ref{IR modelling} we model the optical and IR light curves, respectively. In Sect.~\ref{discussions} we discuss the SN and CSM properties, as well as the mass-loss history and possible progenitor channels. Section~\ref{summary} summarises the paper.

\section{Observations and data reduction}\label{sec:Obs_red}

\subsection{Discovery and classification}

\begin{figure}
\centering
\resizebox{1\hsize}{!}{\includegraphics{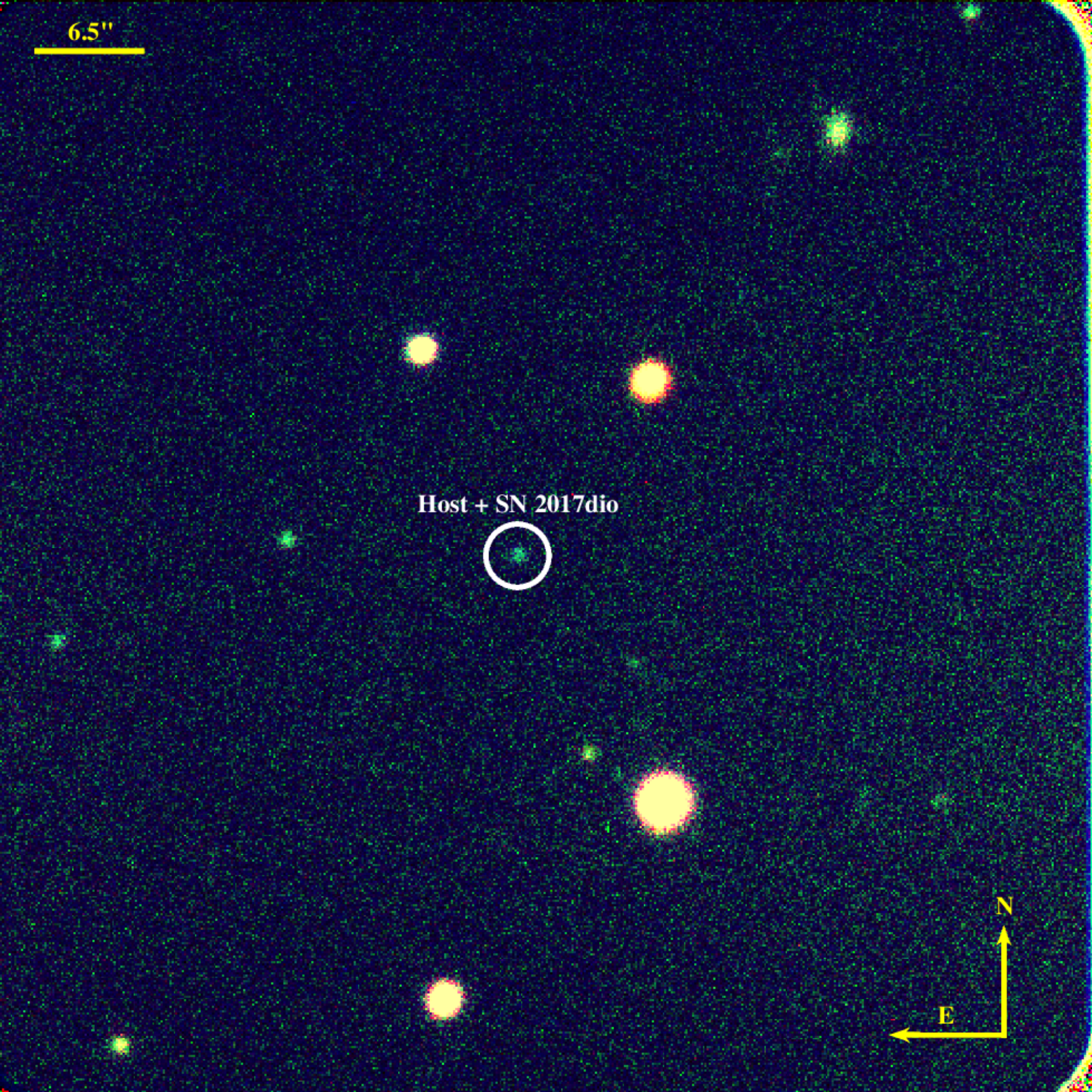}}
\caption{\textit{BVr} composition of SN~2017dio and its host galaxy. Data were taken on 22 October 2017 with the NOT using the ALFOSC instrument.}
\label{fig:hostgalaxy}
\end{figure}

SN~2017dio (ATLAS17evm) was discovered with the Asteroid Terrestrial-impact Last Alert System (ATLAS; \citealt{Tonry18, Smith20}) at RA=11:36:27.770 and Dec=+18:17:47.46 (J2000) in the host galaxy SDSS J113627.76+181747.3. It was first identified on 26 April 2017 (MJD=57869) in the $c$ (cyan) band (5$\sigma$ limit; \citealt{disc_report}). Further analysis using the ATLAS forced photometry server \citep{atlas_server} suggests an earlier detection date. The last non-detection at a 3$\sigma$ level occurred on 17 April 2017 (MJD=57860.39), while the first detection was found to be on 20 April 2017 (MJD=57863.51). We estimated the explosion epoch to be the mid-point between these dates, approximately on 19 April 2017 (MJD=57861.95$\pm$1.56). We used MJD=57862.0 as the explosion date. We refer phases to this epoch, if not otherwise indicated. A \textit{BVr} composition image of SN~2017dio and its host galaxy are shown in Fig. \ref{fig:hostgalaxy}.

\subsection{Photometry}

The light curves of SN~2017dio from the initial discovery to approximately 100 days after the explosion were presented in \citetalias{K18}. In this work, we present new optical and IR observations and analysis of SN~2017dio. The SN was photometrically monitored for nearly 400 days in optical bands and observed at four epochs in the near-infrared (NIR) using various facilities. It was also observed with the Wide-field Infrared Survey Explorer (WISE) as part of the Near-Earth Object Wide-field Infrared Survey
Explorer (NEOWISE) reactivation mission (\citealt{neowise}) in the mid-infrared (MIR) using the \textit{W1} (3.6 $\mu$m) and \textit{W2} (4.6 $\mu$m) filters. The NEOWISE photometry is from co-added images \citep{Lang14,WISE2} that have been template subtracted with High Order Transform of PSF ANd Template Subtraction (HOTPANTS; \citealt{hotpants}). Most optical observations beyond 100 days were obtained by ATLAS in the orange band and by the Liverpool Telescope (LT; \citealt{LT}) using the IO:O instrument in the \textit{BgVriz} bands. The Nordic Optical Telescope (NOT) contributed observations in the \textit{BgVriz} bands using ALFOSC, as well as all four epochs in NIR (\textit{JHK} bands) using NOTCam. All NOT observations were obtained through the NOT Unbiased Transient Survey 2 (NUTS2\footnote{\url{https://nuts.sn.ie}}). Additional \textit{V}-band measurements were obtained by Catalina Sky Survey (CSS; \citealt{CRTS}).
For completeness, we also included data in the $w_{P1}$ band from the Panoramic Survey Telescope and Rapid Response System (Pan-STARRS; \citealt{panstarrs}). 
ATLAS data were retrieved from the ATLAS forced photometry server, with photometric calibration based on the ATLAS Refcat2 catalogue \citep{refcat}. For LT observations, photometric measurements were performed using The AUTOmated Photometry Of Transients (\textsc{Autophot}; \citealt{autophot}), employing point-spread function (PSF) fitting photometry and calibrated using the Sloan Digital Sky Survey (SDSS) catalogue \citep{SDSS} for \textit{griz} bands and the AAVSO Photometric All-Sky Survey (APASS; \citealt{apass}) for the \textit{BV} bands. \textsc{Autophot} was also used for the NOT optical data and three epochs of NIR photometry. NIR calibration was based on the Two Micron All-Sky Survey (2MASS) catalogue \citep{2mass}. New optical and NIR photometry of SN~2017dio are presented in Tables~\ref{tab:atlas}, \ref{tab:Pan-STARRS}, \ref{tab:optical_1}, and \ref{tab:NIR_phot}.

\subsection{Spectroscopy}

The first nine spectra of SN~2017dio were presented in \citetalias{K18}; the reader is referred to that paper for details on the reduction procedures. The final three spectra were obtained with the NOT as part of the NUTS2 collaboration, using the ALFOSC instrument. The last two epochs were reduced and calibrated (in flux and telluric correction) using spectrophotometric standard stars observed during the same nights and processed with the \textsc{alfoscgui}\footnote{\textsc{alfoscgui} is a variant of \textsc{foscgui}, which is a graphical user interface aimed at extracting SN spectroscopy and photometry obtained with FOSC-like instruments. It was developed by E. Cappellaro. A package description can be found at \url{sngroup.oapd.inaf.it/foscgui.html.}} pipeline. 
The spectrum taken on the 1st December 2017 was flux calibrated with a standard star observed on 6 December 2017. All spectra are flux calibrated to match \textit{g}-band photometry and corrected for galactic reddening (see Sect.~\ref{host_gal}). The details of the instruments used for the spectroscopic observations are listed in Table \ref{tab:spec}. All spectra are made publicly available through the WISeREP archive \citep{wiserep}.

\section{Host galaxy}
\label{host_gal}

As reported by \citetalias{K18}, the host galaxy of SN~2017dio is SDSS J113627.76+181747.3, located at a redshift of $z=0.037$, corresponding to a distance modulus of 36.0 mag. The galaxy is a dwarf galaxy, with photometric measurements of $u=23.36\pm1.09$, $g=21.12\pm0.06$, $r=20.67\pm0.06$, $i=20.55\pm0.09$, $z=20.12\pm0.23$ \cite[SDSS DR12;][]{SDSS_DR12}, $J=19.95\pm0.2$, $H=19.24\pm0.16$, $K >19.39$  (photometry taken with the NOT on 22 April 2024), $W1=18.61\pm0.57$, $W2 >18.03$ (from templates used in host galaxy subtraction; \citealt{Lang14,WISE2}). These magnitudes suggest that the host is a low-luminosity galaxy, with an absolute magnitude of $M_g=-15.1$ mag. The Galactic reddening along the line of sight is $E(B-V)=0.028$ mag \citep{extinction}, and the host galaxy extinction component is negligible, as no narrow Na~I~D absorption is detected at the host redshift.

\begin{table}
\small
\caption{Median host galaxy parameters derived from our SED fitting with 16th and 86th percentiles subtracted from the median value from \textsc{prospector} and the median values for Type IIn, Type Ic, and Type Ic-BL host galaxies.} 
\label{tab:host_gal} 
\centering 
\begin{tabular}{lllll} 
\hline\hline 
&log(M$_*$) & log(SFR) & log(sSFR) & log(Z)\\
&($M_{\odot}$)&$(M_{\odot}\mathrm{yr}^{-1})$& ($\mathrm{yr}^{-1}$) &($Z/Z_\odot$)\\
\hline 
\\
SN~2017dio&$8.03^{+0.28}_{-0.33}$ & $-4.70_{-0.08}^{+0.04}$ & $-13.02^{+0,04}_{-0.05}$ & $0.75^{+0.49}_{-0.52}$\\
Type IIn&$9.63^{+0.12}_{-0.12}$ & $-0.09_{-0.07}^{+0.07}$ & $-9.70^{+0.07}_{-0.07}$ & $-1.16^{+0.03}_{-0.03}$\\
Type Ic&$9.88^{+0.14}_{-0.14}$ & $0.02_{-0.12}^{+0.11}$ & $-9.79^{+0.07}_{-0.07}$ & $-1.39^{+0.03}_{-0.03}$\\
Type Ic-BL&$8.98^{+0.16}_{-0.16}$ & $-0.33_{-0.16}^{+0.15}$ & $-9.46^{+0.14}_{-0.13}$ & $-1.11^{+0.05}_{-0.05}$\\
\\
\hline
\end{tabular}
\tablefoot{The values for the Type IIn, Type Ic and Type Ic-BL host galaxies are from \cite{host_galaxies}.}
\end{table}

To derive the physical properties of the host galaxy, we modelled its photometric spectral energy distribution (SED) using \textsc{prospector}\footnote{\url{https://prospect.readthedocs.io/en/latest/}} \citep{Johnson21-prosp, Leja17}. \textsc{prospector} fits stellar population models via Monte Carlo sampling of the posterior distributions using \textsc{emcee} \citep{Foreman-Mackey13}, with composite stellar populations generated through FSPS \citep{Conroy09, Conroy10} and the \textsc{miles} spectral library \citep{Falcon-Barroso11-miles}. Following the procedure described in \citet{GG25}, we fitted the $ugrizJHK$ photometry of the host galaxy. The resulting host galaxy parameters from our SED modelling are listed in Table \ref{tab:host_gal}. The corresponding host galaxy SED and the corner plot of the fitted parameters are shown in Figs. \ref{fig:SED_gal} and \ref{fig:mcmc_gal}, respectively. We note that these are rough estimates as they were not accompanied by spectral analysis. In Table \ref{tab:host_gal} we also list the median values for Type IIn, Type Ic, and Type Ic-BL host galaxies reported in \cite{host_galaxies}.

Comparing the stellar mass with a sample of host galaxies for Type IIn, Ic, and Ic-BL \citep{host_galaxies}, we find that the host galaxy of SN~2017dio has a lower mass, but also higher metallicity than the median of the host galaxies of these types. The star formation rate (SFR) is the lowest found in any of the host galaxies, with only a few of Type IIn and Type Ic host galaxies coming close.

\section{Photometric properties} 
\label{sec:photometric_properties}

\subsection{Light curves}

\begin{figure}
\centering
\resizebox{1\hsize}{!}{\includegraphics{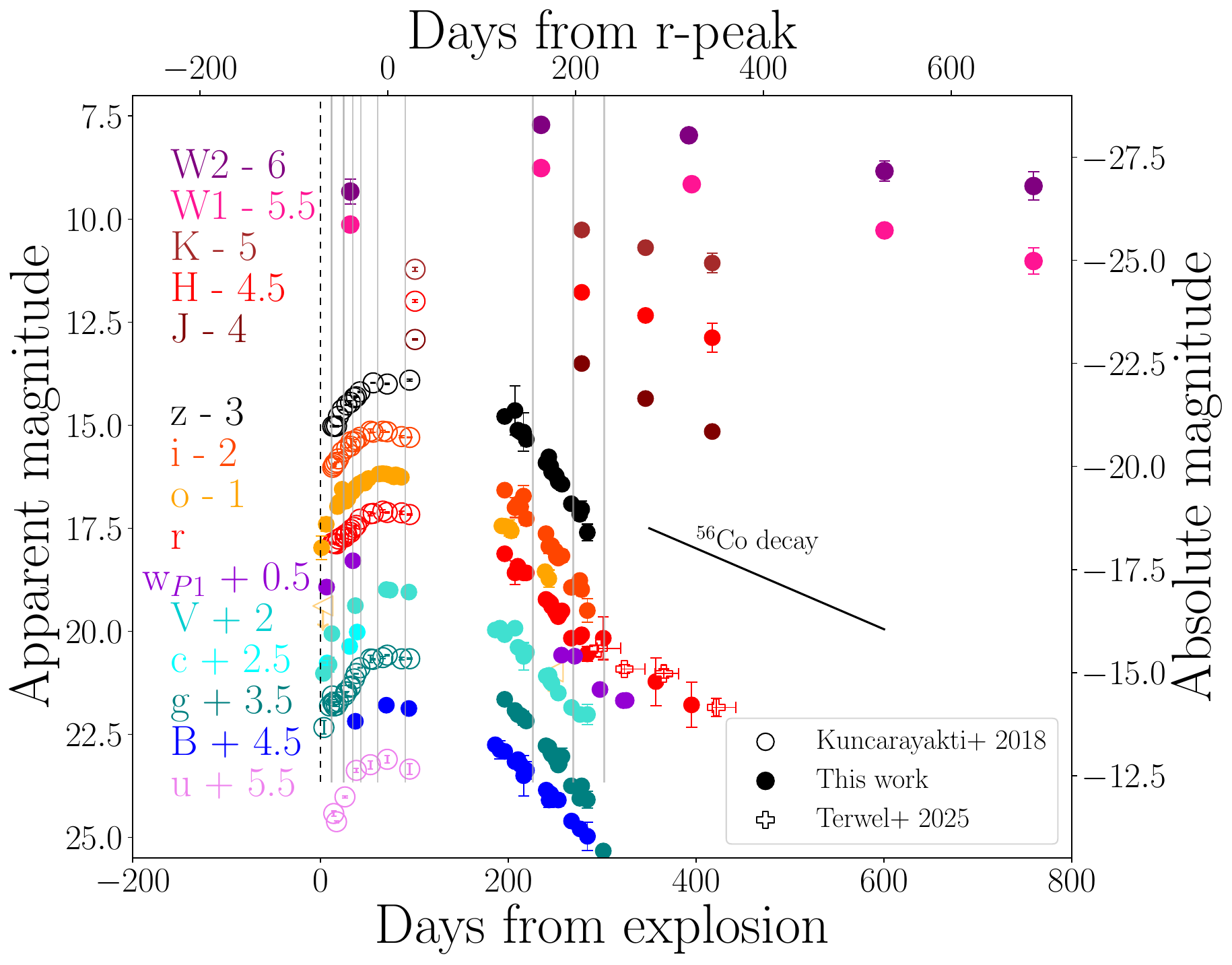}}
\caption{Light curves of SN~2017dio. The epochs are days from explosion (bottom x-axis) and days from $r$-band peak (top x-axis). The epochs of spectra are marked with grey lines, and the explosion epoch is marked with a dashed black line. The data from \citetalias{K18} are marked with hollow circles, from this paper with coloured circles, and from \citet[pre-ZTF stacked data]{Terwel} with hollow plus signs. The light-curve decline following the $^{56}$Co decay is marked with a black line.}
\label{fig:LC}
\end{figure}

The complete light curves of SN~2017dio are shown in Fig. \ref{fig:LC}. The early light curves of SN~2017dio were previously analysed by \citetalias{K18}, who reported a bump in the light curve that occurs during the rise to the peak. Their analysis suggested that the bump was powered by $^{56}$Ni decay, while the subsequent evolution, up to approximately 100 days, was powered by the interaction between the ejecta and the CSM. This interpretation is supported by the spectral evolution, which becomes increasingly dominated by interaction signatures, as well as by the unusually long rise to peak brightness followed by a decline steeper than that expected from the $^{56}$Co decay (0.98 mag per 100 days; \citealt{Woosley89}).

Rise times to the peak brightness in the optical bands were estimated using a Gaussian process (\citealt{gp}) by fitting the data from the explosion to $\approx 100$ days using a Matérn 3/2-Kernel, excluding the \textit{c} and $w_{P1}$ bands due to sparse data coverage. Peak magnitude errors are determined by the standard deviation provided by a Gaussian process, and the errors of the date are calculated as the difference between the estimated and observed peak times (including the error in the explosion epoch). Among the optical filters, we found that the rise times are fastest in the $o$ band and slowest in the $z$ band, with rise times ranging from $\sim66$ to 78. Rise times for the IR bands were not calculated due to insufficient data; however, it is evident that the IR peaks occur later than those in the optical, either during or after the observational gap. The optical light curves exhibit a late-time decline rate of approximately 0.034-0.04 mag day$^{-1}$, derived by fitting a line to the late-time data.
Table \ref{tab:peak_time} summarises the absolute peak magnitudes and corresponding rest-frame rise times.

\begin{table}
\centering
\caption{Peak absolute magnitudes (Peak abs. mag), rise times, and decline rates for SN~2017dio in optical bands.}
\label{tab:peak_time}
\begin{tabular}{ccccllllllll}
\hline\hline
Band & Peak abs. mag & Rise time & Decline rate\\
     & (mag) & (d) & (mag/day)\\
\hline
$u$ & $-18.65\pm0.10$ & $74\pm2$ & $\cdots$\\
$B$ & $-18.72\pm0.18$ & $73\pm2$ & $\cdots$\\
$g$ & $-18.96\pm0.02$ & $69\pm1$ & 0.038\\
$V$ & $-18.52\pm0.01$ & $67\pm2$ & $\cdots$\\
$r$ & $-18.94\pm0.13$ & $72\pm2$ & 0.034\\
$o$ & $-18.83\pm0.02$ & $66\pm1$ & $\cdots$\\
$i$ & $-18.86\pm0.03$ & $70\pm1$ & 0.039\\
$z$ & $-19.18\pm0.21$ & $78\pm4$ & 0.040\\
\hline
\end{tabular}
\tablefoot{The uncertainties in Peak abs. mag come from the fit and the uncertainties in rise times are to the closest observed epochs.}
\end{table}

\subsection{Infrared excess}\label{IR-excess}

\begin{figure}
\resizebox{\hsize}{!}{\includegraphics{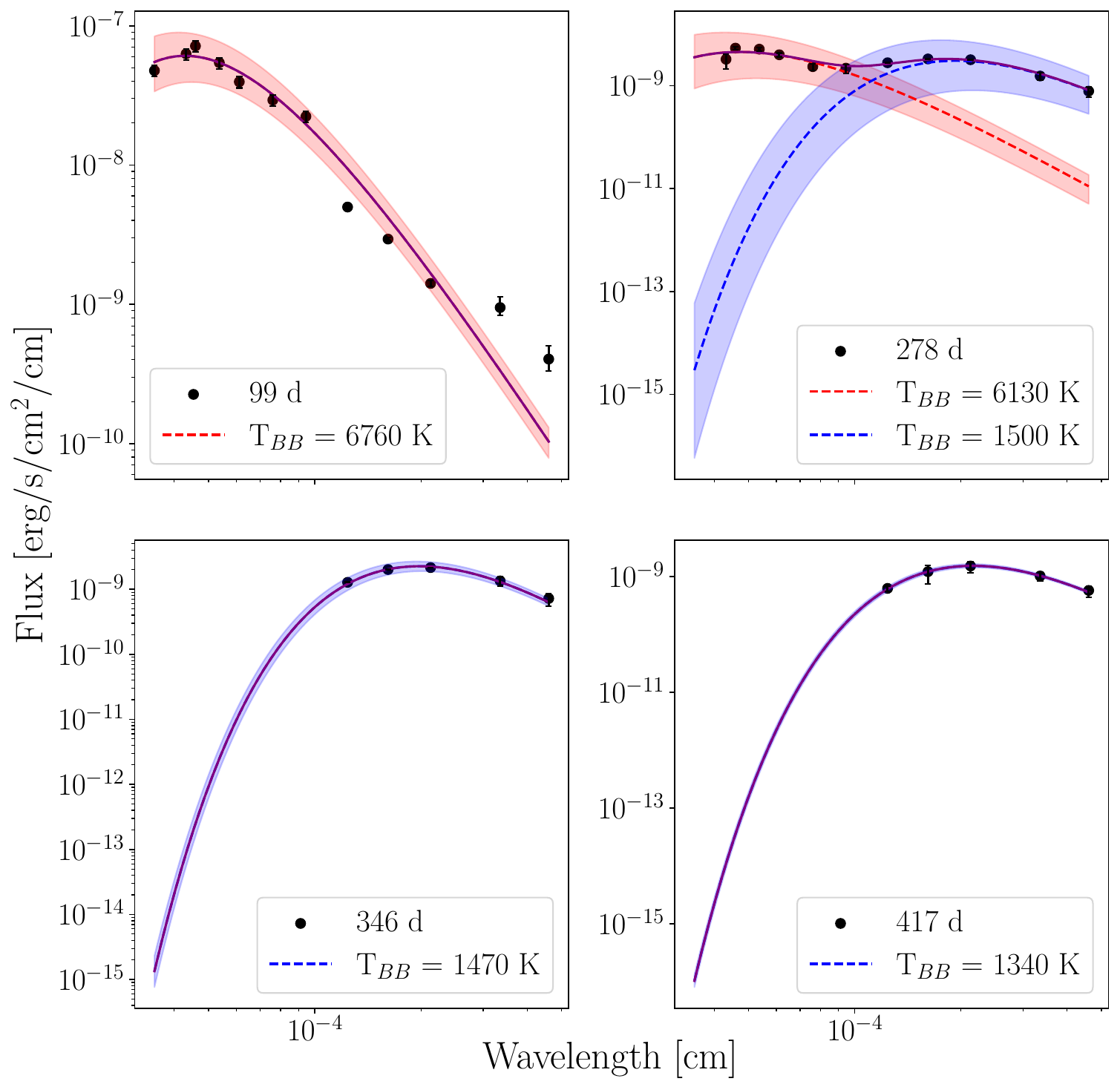}}
\caption{SEDs of SN~2017dio at the four NIR epochs. All filters, except for NIR, have been interpolated to match the epochs. The black points show the flux during each epoch and are given in days from the explosion. The dashed red line shows the hot BB component contributing in the optical bands, while the dashed blue line shows the warm BB component dominating in the IR. The filled*in areas are 1$\sigma$ errors for the fits. The solid purple line shows the best-fit two-component BB.}
\label{fig:sed}
\end{figure}

The IR light curves of SN~2017dio exhibit a delayed peak and a significantly slower decline compared to the optical bands. To investigate the presence of an IR excess, we constructed SEDs for the available epochs with NIR photometry (Fig. \ref{fig:sed}). At early times ($\sim99$~days), the SED shows no significant IR excess, displaying a single hot blackbody (BB) function consistent with thermal emission from the SN itself. However, by the next epoch ($\sim278$~days), while the direct emission from the SN (traced by the optical bands) becomes weaker, a second thermal component emerges in the IR bands, with a comparable strength to the optical component, dominating the NIR. At $\sim346$ and $\sim417$~days, the original SN BB has faded substantially, yet the IR BB component remains prominent. This persistent emission at late times provides strong evidence for a sustained IR excess, likely arising from pre-existing  or newly formed dust. 

To quantify the properties of the IR excess, we fitted two BB functions to the SEDs: one representing the direct SN emission and the other accounting for the IR excess. This way, we subtracted the SN component from the IR bands, which left us with fluxes only from the IR excess emission. At 99 days, reliable magnitudes for the IR excess could not be extracted for ${JHK}$ bands as they are all fainter than the best fit BB. This suggests that a significant IR excess was not present at this epoch. By 278 days, we estimate a BB temperature of 1500$\pm$150 K for the IR component, which is consistent with emission from hot dust. This temperature is very close to the evaporation temperature of silicate dust \citep[$\mathrm{T_{evp} \approx 1500 K}$;][]{dust_evap}.

\subsection{Pseudo-bolometric light curve}

\begin{figure}
\resizebox{\hsize}{!}{\includegraphics{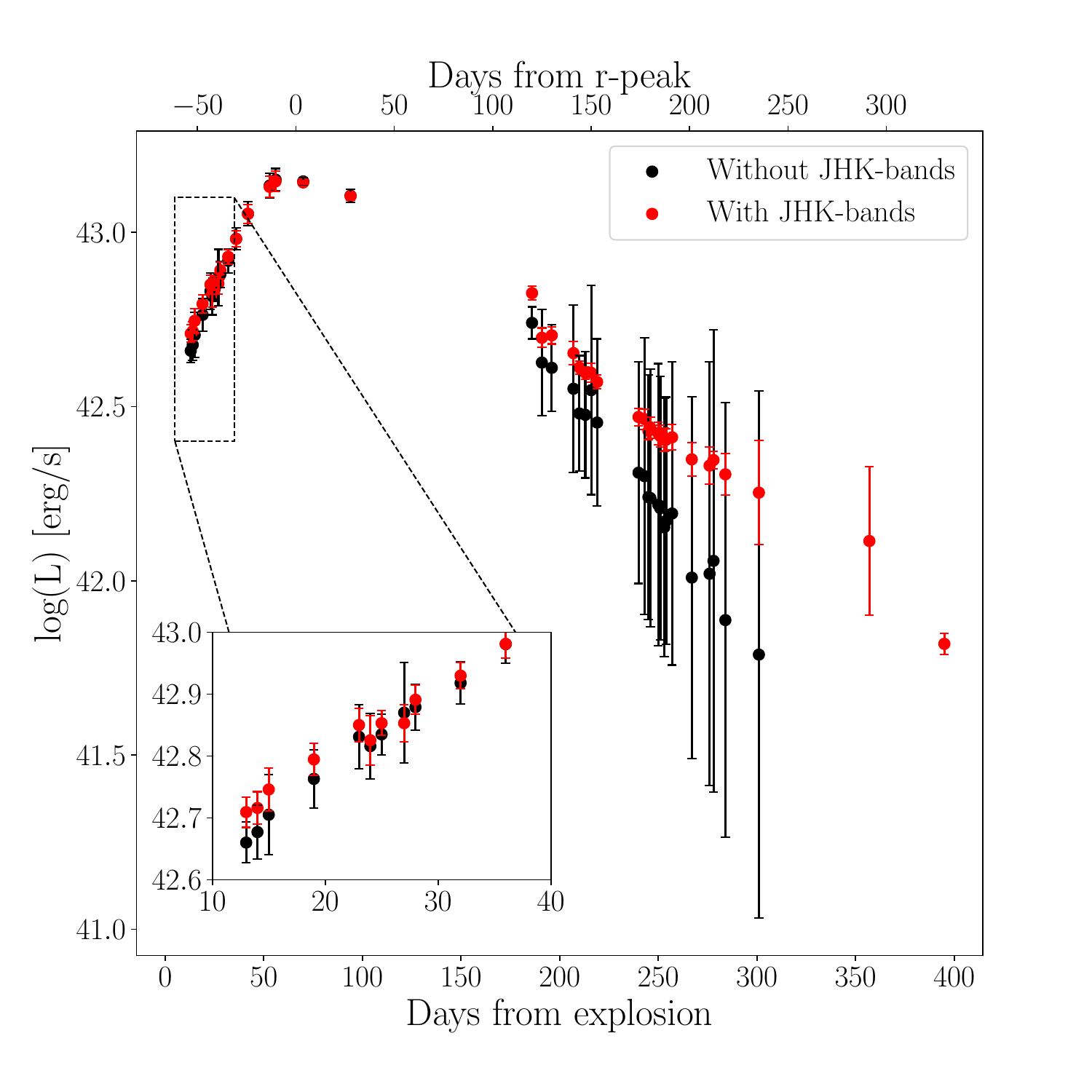}}
\caption{\textit{uBgVroiz} pseudo-bolometric light curves of SN~2017dio with $JHK$ bands (red) and without $JHK$ bands (black). The y-axis is the logarithmic luminosity (L), and the x-axis is in days from r-peak (\textit{top}) and days from the explosion (\textit{bottom}). An inset showing day 10 to day 40 post-explosion is displayed in the left-lower corner to show the bump in the light curve. }
  \label{fig:bol}
\end{figure}

\begin{figure}
\resizebox{\hsize}{!}{\includegraphics{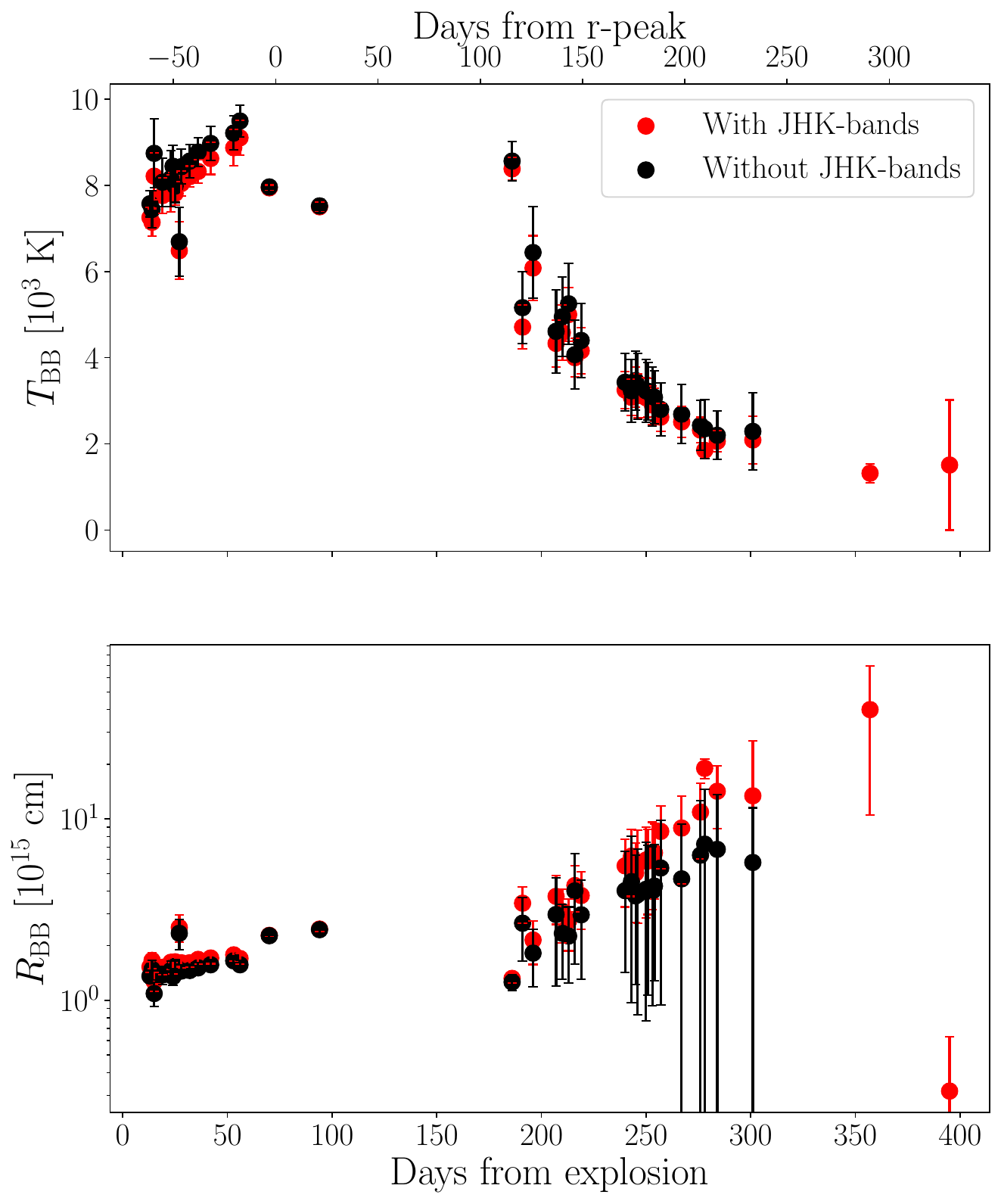}}
\caption{BB temperature (\textit{top}) and BB radius (\textit{bottom}) of SN~2017dio. Red points represent values derived using both optical and NIR bands, while black points show the results obtained when the NIR bands are excluded.}
\label{fig:bb_t_r}
\end{figure}

Two pseudo-bolometric light curves of SN~2017dio were constructed with \textsc{superbol} \citep{superbol}, one based on the \textit{uBgVroiz} bands and the other also including the \textit{JHK} bands. These bolometric light curves are shown in Fig. \ref{fig:bol}. The bolometric light curve without the \textit{JHK} bands shows the SN behaviour without the influence of the significant IR excess at late times. Similarly, we also constructed the BB temperature and radius of SN~2017dio using \textsc{superbol} with and without the NIR bands. These can be seen in Fig. \ref{fig:bb_t_r}.

The bolometric light curves show similar evolution during early phases ($<100$~days) and manage to recreate the bump during the rise. The bump has a luminosity of $\mathrm{L_{bump}} = 10^{42.83 \pm0.05}$ erg s$^{-1}$ and is at 23 days. The main peak, has a luminosity of $\mathrm{L_{peak}}= 10^{43.15 \pm 0.03}$ erg s$^{-1}$ and is at 70 days (luminosities from bolometric light curve including all bands). The timings of both the bump and the peak are consistent with the optical bands. At later times, the light curve that excludes the \textit{JHK} contribution declines nearly twice as fast (decline rate $\approx 0.02$ mag day$^{-1}$) as the one including NIR bands (decline rate  $\approx 0.012$ mag day$^{-1}$), highlighting the increasing influence of the IR excess. While the BB temperature derived in the two cases remains similar, the inferred BB radius is systematically smaller when the NIR bands are omitted. This is consistent with the lower luminosity measured at these epochs.

We estimated the $\mathrm{{^{56}Ni}}$ synthesised in the SN explosion mass using Arnett's rule (\citealt{arnett1}, \citeyear{arnett2}), as follows:

\begin{equation}
    M_{Ni} = \frac{L_{peak}}{(\epsilon_{Ni}-\epsilon_{Co})e^{-t_p/t_{Ni}}+\epsilon_{Co}e^{-t_p/t_{Co}}},
\end{equation}

\noindent where $L_{peak}$ is the peak bolometric luminosity, $\epsilon_{Ni} = 3.9 \times 10^{10}$ erg g$^{-1}$ s$^{-1}$ and $\epsilon_{Co} = 6.8 \times 10^9 $ erg g$^{-1}$ s$^{-1}$ are the specific heating rates of Ni and Co decay, $t_{Ni}$ = 8.8 days and $t_{Co}$ = 111.3 days are the decay scales of Ni and Co and $t_p$ is the rise time. Since the main peak seems to be primarily powered by interaction, we instead adopted the bolometric luminosity of the bump for our estimate, which is mainly powered by $^{56}$Ni decay. In this case, we used $L_{bump} = 10^{42.83}$ erg s$^{-1}$ and $t_p = 23$ days. This gives us an estimated $\mathrm{{^{56}Ni}}$ mass of M($\mathrm{{^{56}Ni}}) \approx 0.43\pm0.05~ M_{\odot}$. We note that this is the upper limit for the $\mathrm{M({^{56}Ni})}$ as Arnett's rule tends to overestimate the $\mathrm{M({^{56}Ni})}$ in some cases when compared with other methods (see e.g. \citealt{nickel_thing2}, \citealt{nickel_thing1} and \citealt{arnett_over}). The high $^{56}$Ni mass could also indicate that the early-time luminosity is affected by interaction as well and is not purely powered by Ni decay as assumed earlier.

The total radiated energy emitted by SN~2017dio was estimated by integrating the bolometric light curve (excluding the \textit{JHK} bands) from 0 to 300 days, giving a value of approximately $2.5 \times 10^{50}$ erg. This should be considered a lower limit as it is derived from the pseudo-bolometric light curve and was integrated only up to 300 days. We estimated the total radiated energy from the IR excess by constructing a pseudo-bolometric light curve using only the IR bands and integrating it from 0 to 400 days. We find that the total radiated energy in the IR bands is $9.6 \times 10^{49}$ erg, which is $38\%$ of the total radiated energy in optical bands. 

\subsection{Photometric comparison}

\begin{figure}
\resizebox{\hsize}{!}{\includegraphics{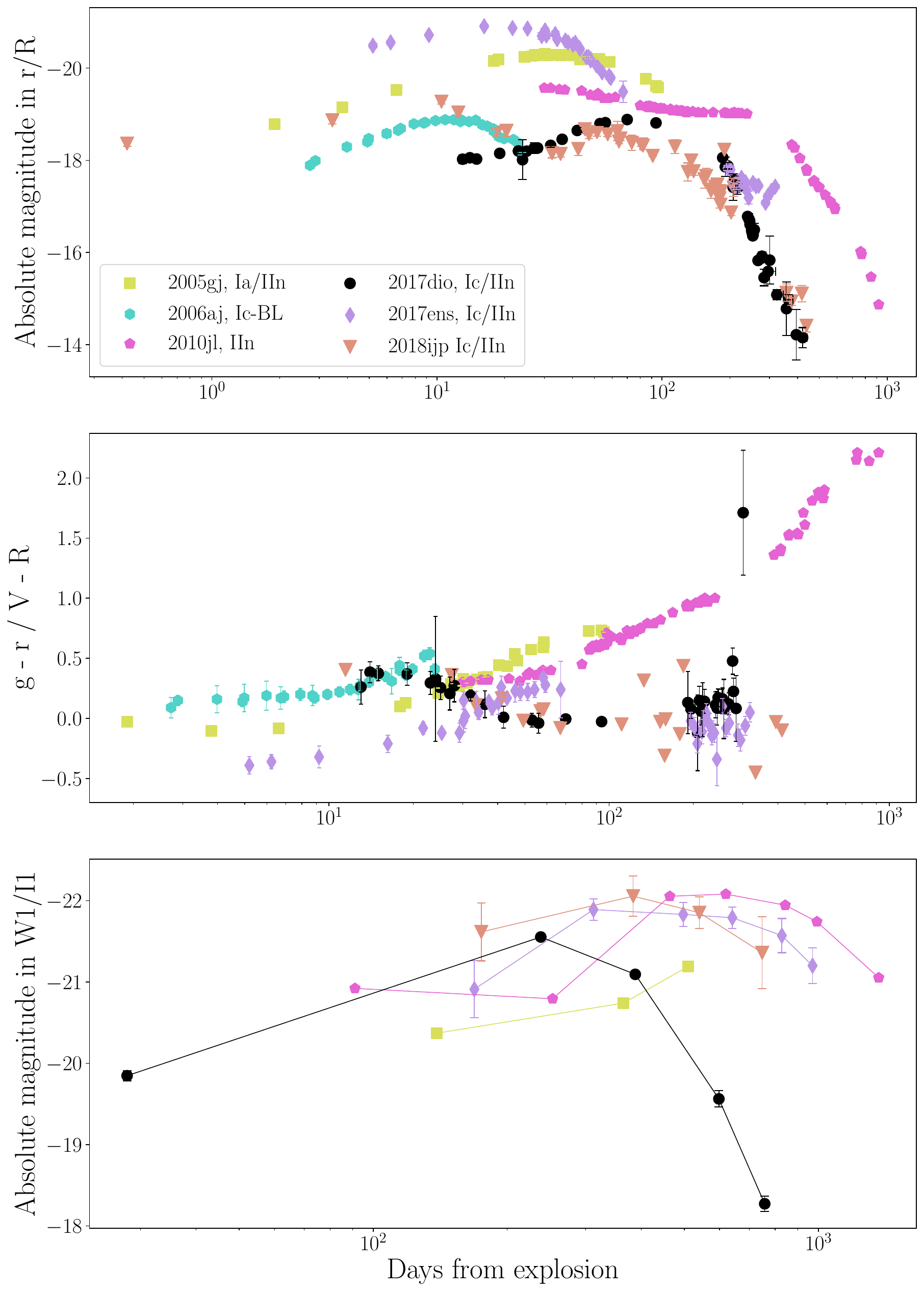}}
\caption{\textit{Top}: $r/R$-band light curve of SN~2017dio (black circles) compared with those of SN~2005gj \cite[green squares;][]{2005gj_phot}, SN~2006aj \cite[blue hexagons;][]{2006aj_phot}, SN~2010jl \cite[red pentagons;][]{2010jl_phot}, SN~2017ens \cite[violet diamonds;][]{2017ens}, and SN~2018ijp \cite[brown triangles;][]{Tartaglia}. \textit{Middle}: $g-r / V-R$ colour evolution of SN~2017dio compared with the same set of comparison SNe. The data for SN~2010jl in both the top and middle panels have been cut to only include data from 450 days from the explosion. \textit{Bottom}: Comparison plot of the $W1$ (3.6 $\mu$m) light curves of SN~2017ens \citep{2017ens_IR}, SN~2017dio, and SN~2018ijp, and the $I1$ (3.6 $\mu$m) light curves of SN~2005gj \citep{2005gj_IR} and SN~2010jl \citep{2010jl_IR}. Markers and colours are the same in all panels.
}
\label{fig:lc_comp}
\end{figure}

Figure \ref{fig:lc_comp} presents the \textit{r/R}-band light curves of SN~2017dio alongside those of a comparison sample. These comparison SNe were selected based on spectral similarities at different phases, identified using \textsc{gelato} \citep{gelato} and \textsc{snid} \citep{snid}. We included SN~2017ens \citep{2017ens} and SN~2018ijp  \citep{Tartaglia} specifically because they exhibit a similar evolutionary transition from a Type Ic to a Type IIn SN. 

The light curve of SN~2017dio diverges noticeably from those of the comparison objects. Since SN~2006aj is a Type Ic-BL SN \citep{Sollerman_2006}, its light curve provides a useful baseline for assessing the underlying SN component in SN~2017dio. Assuming a similar intrinsic evolution, it becomes clear that the dominant source powering SN~2017dio is the interaction between the ejecta and the surrounding CSM. Among the comparison objects, SN~2005gj displays the closest match in decline rate right after peak, differing by only $\sim0.004$ mag/day. SN~2017ens declines more rapidly immediately after the peak, but once SN~2017dio begins to fade (at around $\sim200$ days), the tail of SN~2017ens declines at a much lower rate. SN~2010jl also shows a substantially shallower decline than SN~2017dio. The light curve of SN~2018ijp shows two peaks, both of them peaking earlier than the bump and peak in SN~2017dio. The first peak of SN~2018ijp is also considerably brighter than the bump in SN~2017dio, while the second peak is slightly dimmer. After the peaks, SN~2017dio declines more rapidly, but during the tail-phase SN~2017dio and SN~2018ijp have similar declines. Interestingly, the light curves of these two SNe have the same duration; neither has any detections after $\sim 400$ days.

\begin{table*}
\small
\caption{Detailed properties of the comparison sample: name, SN type, redshift, explosion epoch, $r$-peak time, and decline rate.}
\label{tab:comp}
\centering 
\begin{tabular}{c c c c c c c}
\hline\hline 
SN & Type & $z$ & Exp. epoch & $r$-peak time & Decline rate & References\\
&&& (MJD)&(d)&(mag/day)&\\
\hline
2005gj & Ia/IIn & 0.0616 & 53637.43& 30 & 0.014&\cite{2005gj_spec1}, \cite{2005gj_phot},$\dagger$\\
2006aj & Ic-BL & 0.033 & 53784* & 11 & 0.079 &\cite{2006aj_spec2}, \cite{2006aj_spec1}, \cite{2006aj_phot} \\
2010jl & IIn & 0.0107 & 55479** & $\sim12$*** & 0.088 & \cite{2010jl_spec1}, \cite{2010jl_spec2}, \cite{2010jl_spec3} \\
2017ens & Ic/IIn & 0.1086 & 57907.8& 10 & 0.036  &\cite{2017ens}\\
2018ijp & Ic/IIn & 0.0848 & 58429.58 & 8 & 0.015 & \cite{Tartaglia}\\
\hline 
\end{tabular}
\tablefoot{
$\dagger$ The spectrum taken at 368 days was obtained from WISeREP archive \citep{wiserep} and was observed by N. Morrell\\
*The explosion epoch for SN~2006aj is considered to be the date of the gamma-ray burst.\\ 
**We adapted the explosion epoch for SN~2010jl from \cite{explosion_date_2010jl}.\\
***Due to there not being \textit{R}-band data for 2010jl at peak time, we estimated the \textit{R}-peak as the middle point of \textit{V}- and \textit{I}-band peaks.
}
\end{table*}

The apparent \textit{$g-r$} colour evolution of SN~2017dio is shown in Fig. \ref{fig:lc_comp} (middle panel). At early times, SN~2017dio closely resembles SN~2006aj. However, as the CSM interaction becomes more dominant, SN~2017dio becomes progressively bluer, unlike the comparison SNe. Apart from a brief initial reddening, SN~2017dio and SN~2005gj are completely different. At later phases, SN~2017dio shows some `bumps' in its colour evolution, somewhat reminiscent of those seen in SN~2017ens. The colour evolution of SN~2018ijp loosely follows that of SN~2017dio, although it appears bumpier. Interestingly, SNe that evolve from Type Ic to Type IIn (SN~2017ens, SN~2018ijp and SN~2017dio) all exhibit a blueward evolution and remain comparatively bluer at late times, while the other type IIn SNe (SN~2005gj and SN~2010jl) do not share similar behaviour.

Among the comparison SNe, SN~2005gj, SN~2010jl, SN~2017ens and SN~2018ijp have IR observations, and all of them exhibit a late-time IR excess. SN~2005gj showed MIR excess at 139 days, attributed to pre-existing dust \citep{2005gj_IR}. SN~2010jl displays an IR excess much earlier (roughly 50 days from discovery), and a re-brightening at $\sim500$ days; the early IR excess has been interpreted as an IR echo, while the re-brightening is attributed to newly formed dust (\citealt{Maeda2013}, \citealt{Gall_2014}, \citealt{2010jl_phot},\citealt{2010jl_IR}). SN~2017ens also shows MIR re-brightening at approximately $\sim500$ days, probably caused by newly formed dust \citep{2017ens_IR}. We report that SN~2018ijp also has an IR excess; the first detection was at $\sim 175$ days. For SN~2017dio, the gap in the data makes the onset of the IR excess difficult to constrain, but it is clearly present in the data after the gap, roughly 200 days from the explosion. This timing most closely resembles that of SN~2005gj and SN~2018ijp. The duration of the IR excess in SN~2017dio is shorter than that seen in the two dust-forming SNe, but similar to that of SN~2018ijp, although it is possible that the excess of SN~2018ijp continues even after the last detection.

\section{Spectroscopic properties} 
\label{sec:spectroscopic_properties}

\subsection{Spectral evolution}

\begin{figure}
\centering
\includegraphics[width=\columnwidth]{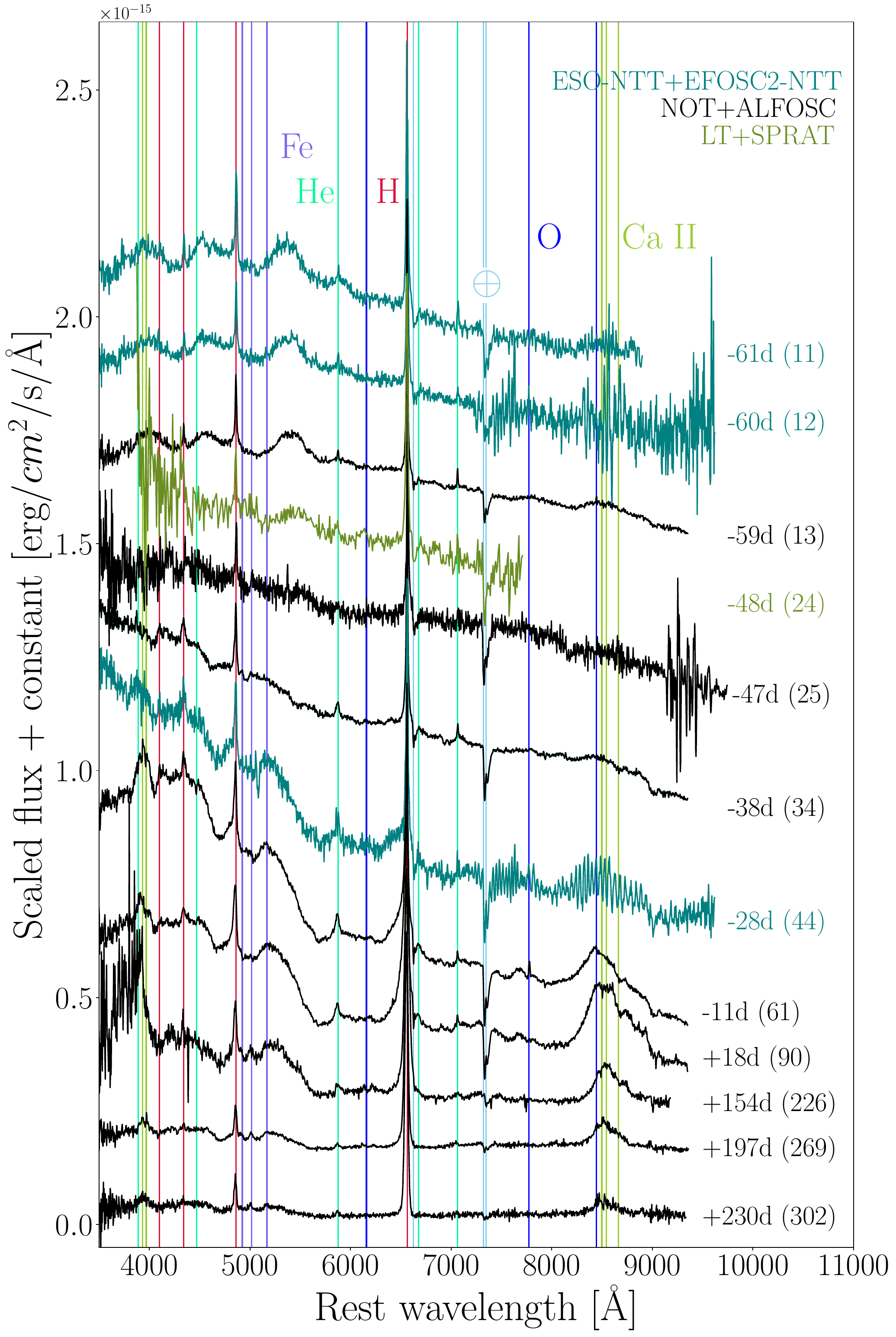}
\caption{Spectral sequence of SN~2017dio. The epochs of the spectra are marked as days from the $r$ peak and days from the explosion (in parentheses). All spectra have been corrected for redshift and scaled to match the H$\alpha$ line intensity for comparison purposes. The vertical lines indicate the rest position of the strongest lines.}
\label{fig:spectral_seq}
\end{figure}

SN~2017dio was classified as a Type Ic SN interacting with hydrogen-rich CSM that later evolves into a Type IIn \citepalias{K18}. Its complete spectral evolution is presented in Fig. \ref{fig:spectral_seq}. The earliest spectra (first three epochs) show broad absorption features characteristic of SNe~Ic, superimposed with prominent narrow hydrogen emission lines (clear evidence of ongoing CSM interaction). At around 25 days, the spectra evolve into a blue, nearly featureless continuum, while the narrow hydrogen emission lines persist. This marks the stage at which CSM interaction becomes the dominant energy source, and the underlying intrinsic SN features are no longer detectable. Notably, this transition coincides with the beginning of the second rise observed in the light curves.

In subsequent epochs, a broad bump emerges in the bluer part of the spectra, similar to the Fe blends identified in SN~2002ic \citep{Deng}. In that event, such features were interpreted as Fe emission arising either from the cool dense shell in the reverse-shocked region or from dense CSM clouds. The spectra also reveal the emergence of Ca~II H\&K absorption in blue, and Ca II NIR triplet in red, potentially blended with O~I $\lambda$8446. These features, both the Fe blends and Ca~II lines, are strongest at around 11 days before the $r$-band peak and 18 days post-peak, after which they gradually fade. In the late-time spectra (the last three epochs), no typical SN Ic nebular lines such as [O~I] and [Ca~II] are detected \citep{Pyykkinen}. Instead, the spectra are dominated by a strong H$\alpha$ emission line. Aside from H, He, Ca, and Fe, no other emission lines are detected in the late phases.

\subsection{Evolution of H$\alpha$}

\begin{figure}
\resizebox{\hsize}{!}{\includegraphics{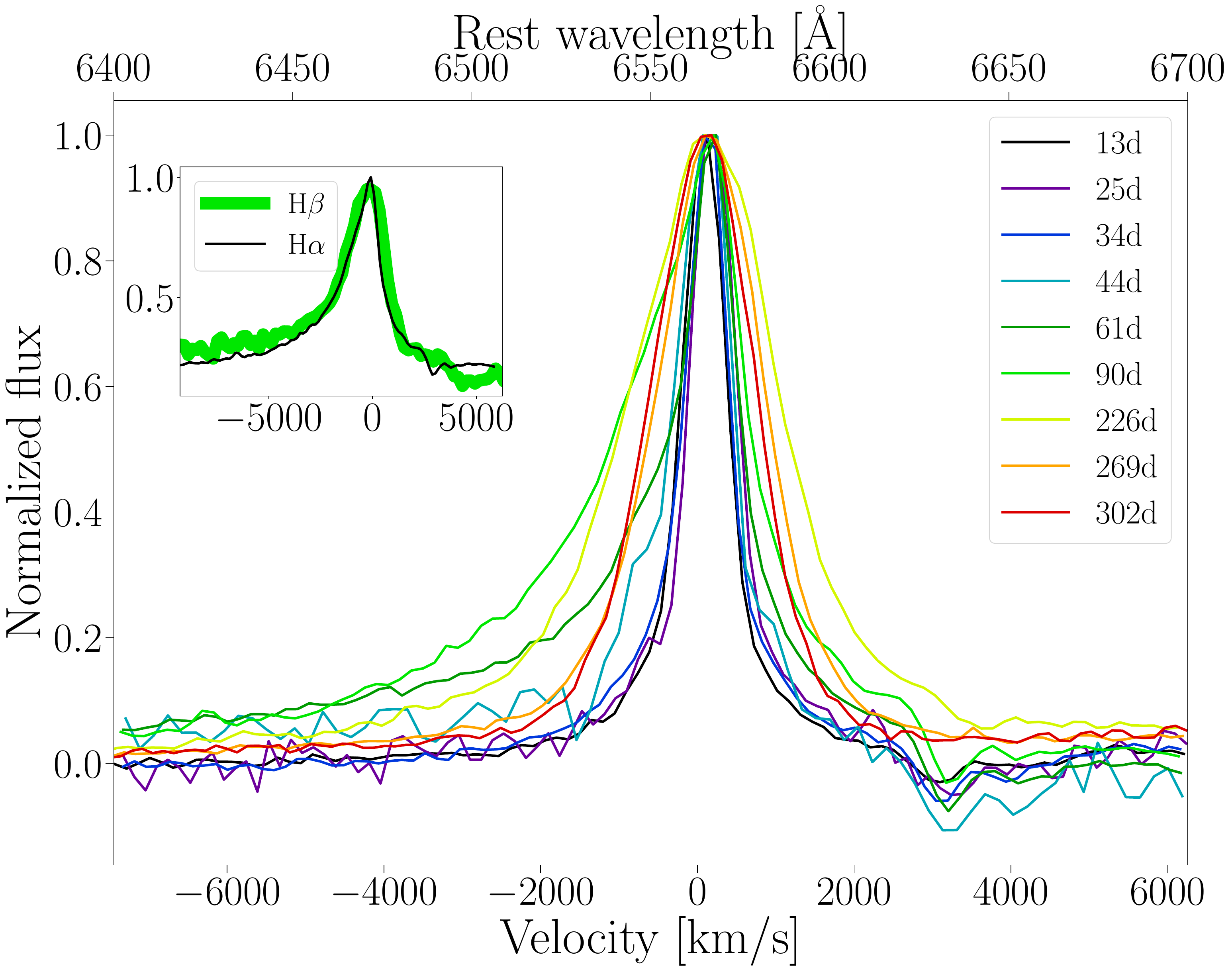}}
\caption{Evolution of the H$\alpha$ line profile of SN~2017dio at different epochs, which are marked as days from the explosion. The H$\alpha$ lines are normalised, and the continuum has been subtracted. The x-axis is shown in velocity space (\textit{bottom}) and in rest wavelength (\textit{top}). The insert shows H$\alpha$ plotted on top of H$\beta$ at 90 days.}
\label{fig:Ha_profile}
\end{figure}

The evolution of the H$\alpha$ line profile from 13 days to 302 days is shown in Fig. \ref{fig:Ha_profile}. For clarity, spectra from similar epochs have been omitted. Initially, H$\alpha$ appears as a narrow feature with Lorentzian wings \citepalias{K18}. A clear change in the line profile becomes apparent at 44 days, when the line starts to broaden, with the broadening continuing until $\sim90$ days. The interaction becomes stronger, as the emitting regions become thicker with stronger electron scattering resulting in an increase in the line widths. It is also observed that the broadening is asymmetric and has an excess in the blue, which could be due to the emergence of the shocked regions \citep{Taddia2020}. We also considered whether the telluric feature at 6650 \AA\ might affect the profile; however, because H$\beta$ displays a similar asymmetry (see the insert in Fig. \ref{fig:Ha_profile}), we conclude that the observed H$\alpha$ asymmetry is intrinsic. 

After 90 days, the H$\alpha$ profile gradually narrows, suggesting that the interaction between the SN ejecta and the CSM is weakening. This behaviour coincides with the decline in the light curves, further indicating that the CSM interaction still dominates the spectra but is steadily diminishing. 
From the profile evolution, we infer that the interaction intensifies around 44 days, reaches its peak at around 90 days, and then gradually weakens. This trend is consistent with the H$\alpha$ luminosity evolution, presented in Fig. \ref{fig:HaHb_ratio}. The H$\alpha$ luminosity remains relatively constant up to 34 days, and then rises until it peaks at 90 days, and subsequently decreases. 
A gap in observations from 90 to 226 days (due to solar conjunction) prevents us from tracking the behaviour during that interval.

Figure \ref{fig:HaHb_ratio} also shows the H$\alpha$/H$\beta$ ratio (the Balmer decrement), which has a theoretical value of 2.86 under the Case B recombination case \citep[e.g.][]{case_b}. Although it is commonly used as a reddening indicator, it is also sensitive to optical depth and therefore to the CSM density. The ratio remains relatively constant at early phases, but rises dramatically at late times, reaching a maximum value of $\sim12$. Such elevated values are frequently observed in SNe IIn \citep{high_B} and are typically attributed to collisional excitations \citep{Branch}.

\subsection{Spectral comparison}

\begin{figure}
\resizebox{\hsize}{!}{\includegraphics{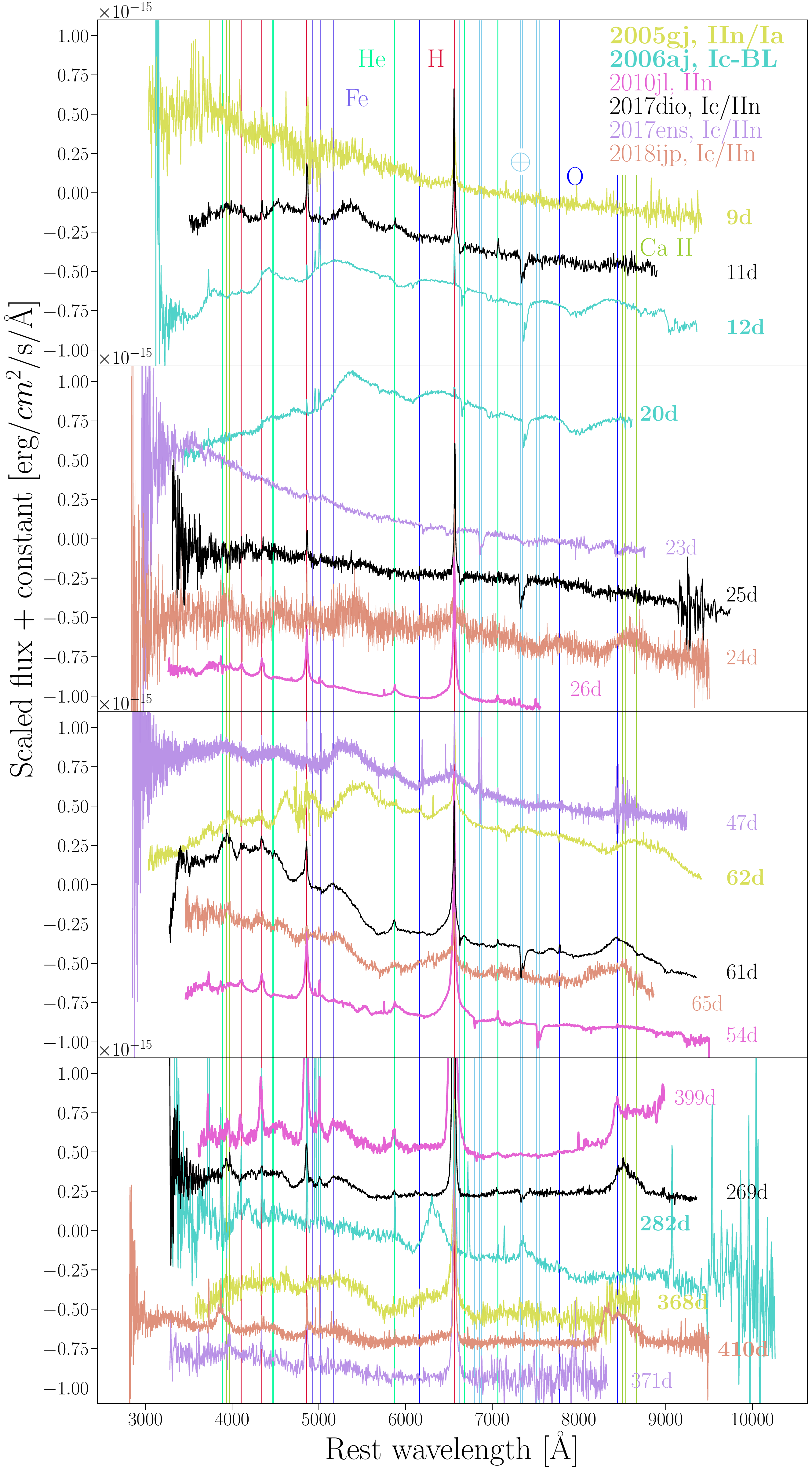}}
\caption{Spectral comparison of SN~2017dio with SN~2005gj (green), SN2006aj (blue), SN~2010jl (red), SN~2017ens (violet), and SN~2018ijp (brown) at four different epochs. The epochs of each spectrum are given in days from the explosion and are marked next to the respective spectra. The spectra have been redshift-corrected, and the flux is scaled.}
\label{fig:spec_comp}
\end{figure}

To better understand the nature of SN~2017dio, we compared its spectra at four different epochs with those of the comparison sample: SNe 2005gj, 2006aj, 2010jl, 2017ens, and 2018ijp. The spectral comparisons are shown in Fig. \ref{fig:spec_comp}. 

At the first epoch, 11 days, SN~2017dio exhibits broad absorption features resembling those of SN~2006aj, though the absorption features for SN~2006aj are significantly broader. This suggests that SN~2017dio had lower ejecta velocities at these early phases. Besides the presence of H$\alpha$ emission, SN~2017dio shows little similarity to SN~2005gj. 

By the second epoch, 25 days after the explosion, its spectrum becomes dominated by signatures of strong interaction. At this point, the continuum turns blue and nearly featureless, closely matching the spectra of SN~2017ens and SN~2010jl. The spectrum no longer matches that of SN~2006aj, indicating the interaction is now the dominant source. 

In the third epoch,  approximately 60 days, SN~2017dio develops a prominent bump in the blue part of the spectrum and shows clear Ca~II NIR triplet emission. At this epoch, the spectrum of SN~2018ijp starts to resemble the spectrum of SN~2017dio, sharing very similar spectral features. The `blue' bump is observed in SN~2010jl, SN~2017ens and SN~2018ijp. However, the lower spectral resolution of SNe 2010ji and 2017ens limits a detailed comparison. All objects show the narrow hydrogen emission lines. Among the comparison SNe, SN~2005gj and SN~2018ijp show a Ca~II NIR triplet emission comparable to that of SN~2017dio. 

The last epoch, approximately 270 days, the spectrum of SN~2017dio is dominated by strong, narrow H$\alpha$ emission, with a much weaker bump and diminished Ca~II NIR triplet compared to earlier epochs. SN~2006aj no longer provides spectral resemblance at this late phase. SN~2005gj continues to show similar features compared to SN~2017dio, such as the remaining bump in the blue part of the spectrum, strong H$\alpha$ emission and possibly the Ca~II NIR triplet. However, the H$\alpha$ profile has a broad component that is not observed in SN~2017dio. SN~2010jl, SN~2017ens and SN~2018ijp also show strong similarity to SN~2017dio, although the spectral coverage of SN~2017ens is limited and noisier, making it difficult to confirm the presence or absence of the Ca~II NIR triplet. SN~2010jl and SN~2018ijp show strong O~I $\lambda8446$ lines, which, if present in SN~2017dio, are blended with the Ca~II NIR triplet. SN~2010jl also shows forbidden [O~III] lines, which are contamination from the host galaxy \citep{2010jl_spec3}.

A direct comparison (Fig. \ref{fig:close_comp}) reveals the spectral resemblance between SN~2017dio, SN~2017ens, and SN~2018ijp. Despite this, the H$\alpha$ to H$\beta$ ratio is lower in SN~2017ens ($\sim 5$) than in SN~2017dio ($\sim12$), which could suggest differences in optical depth or ionisation conditions between the two events. On the other hand, SN~2018ijp ($\sim17$) displays a H$\alpha$ to H$\beta$ ratio greater than SN~2017dio.

\section{Light-curve modelling} 
\label{LC_modelling}

\begin{figure*}
\centering
\begin{subfigure}{0.5\textwidth}
\centering
\includegraphics[height=2.5in]{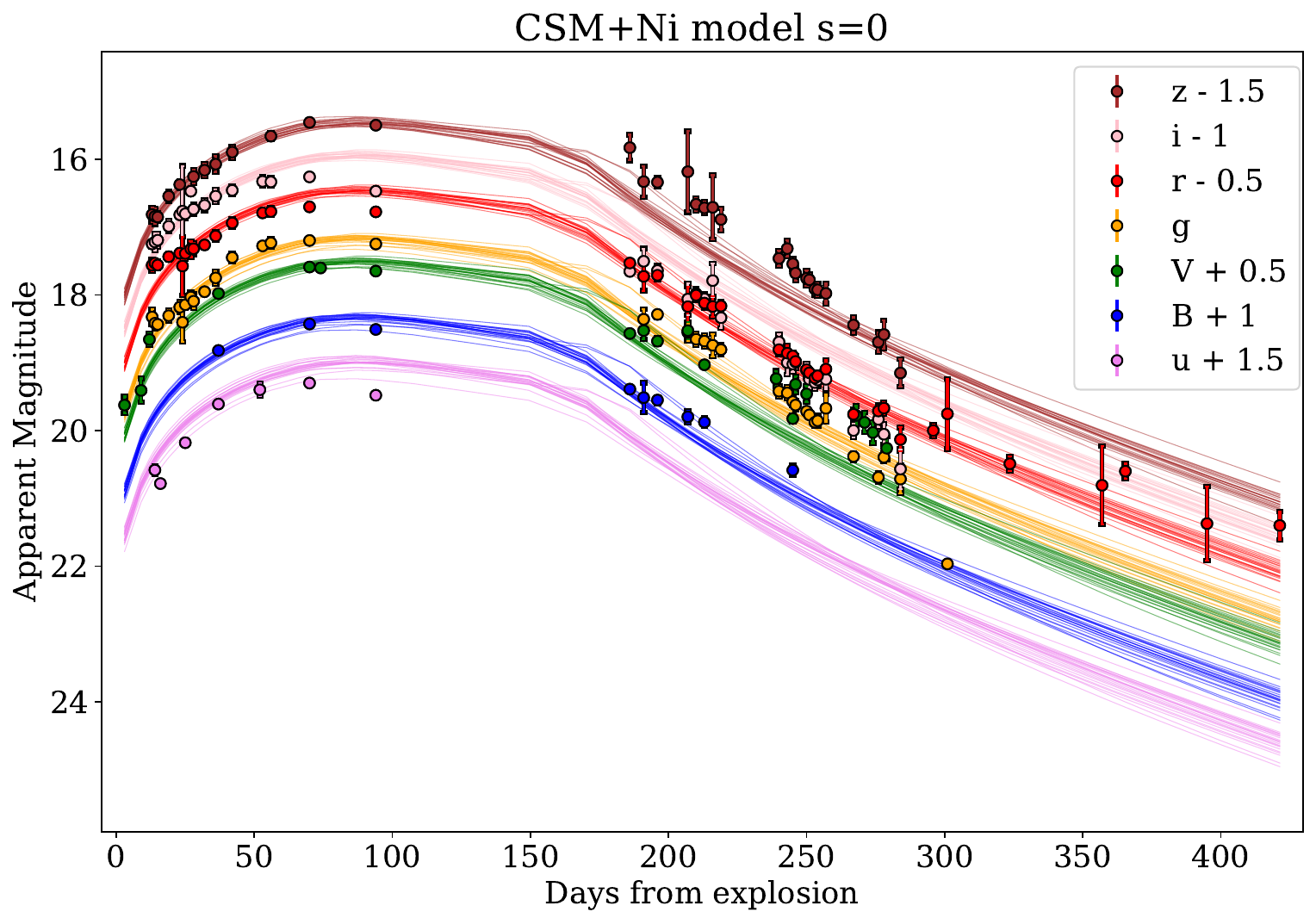}
\end{subfigure}%
~ 
\begin{subfigure}{0.5\textwidth}
\centering
\includegraphics[height=2.5in]{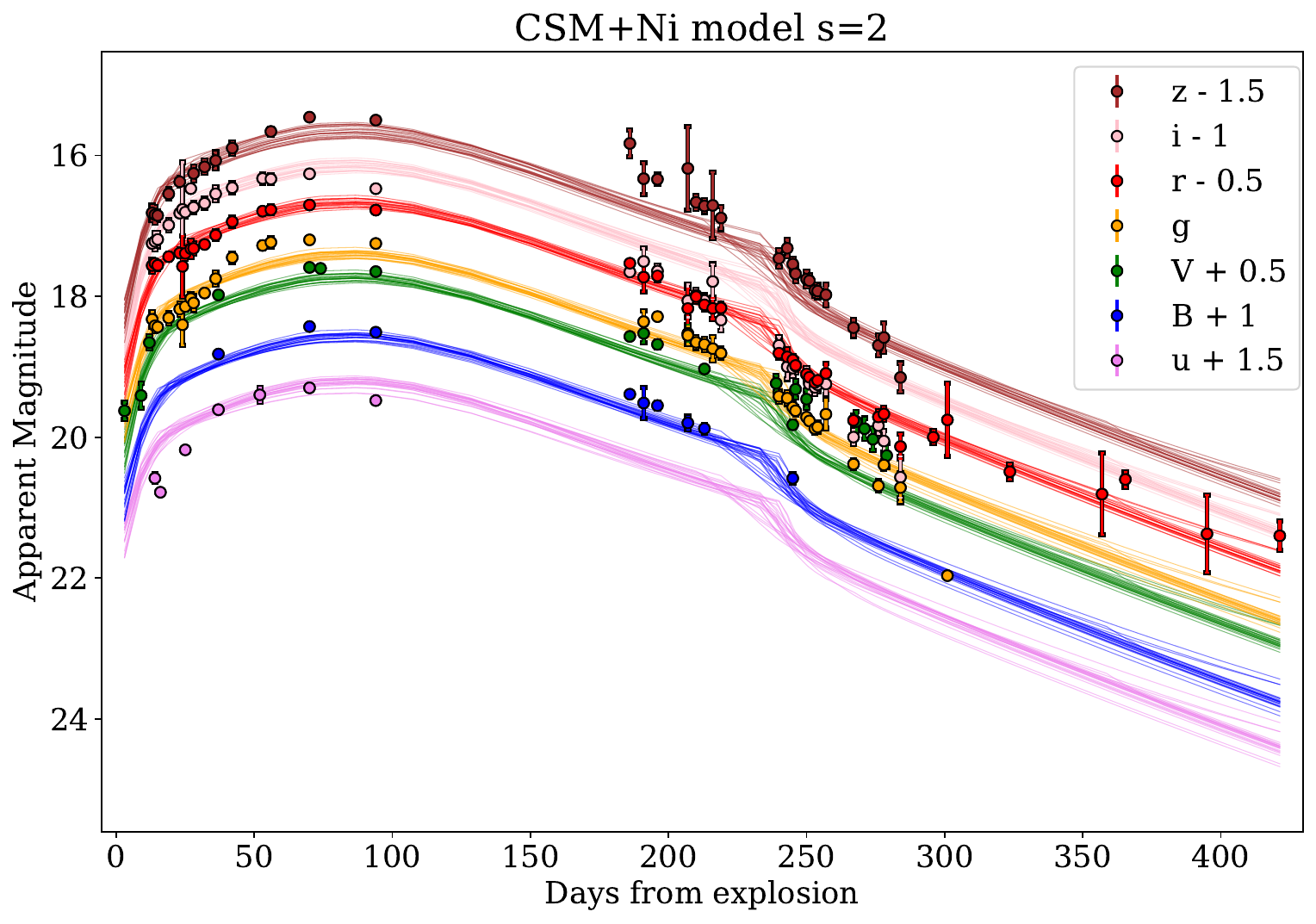}
\end{subfigure}
\caption{Ni+CSM model fitted to the optical photometry of SN~2017dio using MOSFiT. \textit{Left}: Fits for the shell-like density structure ($s=0$). \textit{Right}: Fits for the wind-like density ($s=2$). Fits produced by a random draw of posteriors are plotted to show the range of the models.}
\label{fig:mosfit}
\end{figure*}

We derived the CSM and SN parameters by fitting the \textit{uBVgriz} light curves using the publicly available code Modular Open Source Fitter for Transients \cite[MOSFiT;][]{mosfit}. Because both the SN emission and CSM interaction contribute to the observed light curves, we adopted the \textit{CSM+Ni} model \citep{Cha_2012} and used two CSM density profiles: a shell ($s=0$) and a wind ($s=2$). The remaining optical bands were excluded due to insufficient coverage, and the IR bands were omitted because the IR excess, present in SN~2017dio, is not included in the model. We employed the \textsc{dynesty} dynamic nested sampling package and ran each model until convergence, requiring approximately 18000 iterations for $s=0$ (score 82) and 18000 iterations for $s=2$ (score 83).

\begin{table}
\small
\caption{MOSFiT parameters, their units, and the inferred median values and their 16th and 84th percentile values for s = 0 and s = 2.} 
\label{tab:mosfit} 
\centering 
\begin{tabular}{l l r r r r}
\hline\hline 
Parameter & Unit & \multicolumn{2}{ c }{Inferred values}\\
 & & s = 0 & s = 2\\
\hline 
log($f_{Ni}$) & \%&$ 0.78^{+0.20}_{-0.23}$ & $0.61^{+0.19}_{-0.22}$ \\
log$(M_{CSM})$ &  $\mathrm{M}_{\odot}$ & $1.00^{+0.07}_{-0.08}$ & $0.84^{+0.25}_{-0.07}$ \\
log$(M_{ej})$  & $\mathrm{M}_{\odot}$ & $0.37^{+0.09}_{-0.05}$& $0.73^{+0.14}_{-0.13}$ \\
log$(R_0)$ &  AU  & $1.30^{+0.84}_{-0.86}$ & $1.32^{+0.62}_{-0.80}$ \\
log($\rho_0$) &  $\mathrm{g}~\mathrm{cm}^{-3}$ & $-13.89^{+0.10}_{-0.07}$ & $-12.18^{+1.52}_{-1.30}$ \\
$v_{ej}$ &  $\mathrm{km}~ \mathrm{s}^{-1}$ & $7341^{+876}_{-857}$ & $5747^{+271}_{-325}$ \\
$t_{exp} $& days & $-9.63^{+0.51}_{-0.27}$ & $-5.97^{+0.98}_{-0.94}$ \\
log($\sigma $)&  mag & $-0.50^{+0.03}_{-0.03}$ & $-0.51^{+0.02}_{-0.02}$ \\
log($\kappa_{\gamma} $)&  $\mathrm{cm}^{2}~\mathrm{g}^{-1}$ & $-1.81^{+0.17}_{-0.13}$ & $-1.90^{+0.11}_{-0.07}$ \\
log($n_H$) &  cm$^{-2}$  & $21.34^{+0.05}_{-0.07}$ & $21.38^{+0.03}_{-0.03}$ \\
log($T_{min}) $ & K & $4.21^{+0.06}_{-0.06}$ & $4.25^{+0.04}_{-0.04}$ \\
\hline 
\end{tabular}
\end{table}

The model includes 11 free parameters. The parameter $f_{Ni}$ represents the $\mathrm{{^{56}Ni}}$ mass fraction in the ejecta, varying between 0.1 $\%$ and 20 $\%$. The CSM opacity $\kappa$ was fixed at 0.34 $\mathrm{cm^2g^{-1}}$. The parameter $n$ is a power-law exponent for the SN ejecta density \citep{Cha_2012} and it is related to the SN progenitor with a typical value of $\sim10$ for type Ib/Ic SNe (\citealt{Matzner}, \citealt{Kasen2010}). We explored a broader range of 7 to 12 in several MOSFiT tests, obtaining $n=9.69$ for SN~2017dio. We fixed $n$ to 10 for the final fits due to the common practice of fixing $n$ usually to 7, 10, or 12.  The inner radius of the CSM, $R_0$, is varied from 1 to 500 AU. The explosion date, t$_{exp}$, was constrained to 0-10 days prior to the discovery date, a range that includes the possible explosion time estimate. The ejecta velocity, $v_{ej}$, was varied between 5000 and 18000 km s$^{-1}$. The CSM density, $\rho_0$, was sampled between $10^{-10}$ and $10^{-15}$ $\mathrm{g~cm^{-3}}$ , consistent with the theoretical values for interacting SNe (e.g. \citealt{Yaron17}). The ejecta mass $M_{ej}$ was ranged from 2 to 25
$\mathrm{M}_{\odot}$. We estimated the $M_{ej}$ with

\begin{equation}
    M_{ej} = \frac{1}{2}\frac{\beta c}{\kappa}v_{ph}t^2_{max},
\end{equation}

\noindent where $\beta$ is 13.8, $t_{max}=23$ days, $\kappa = 0.1 \mathrm{cm}^2\mathrm{g}^{-1}$ and $v_{ph}$ ranges from 6000 to $18000\,\rm{km}\,\rm{s^{-1}}$. The velocities were chosen by fitting a line to the Fe II line in early-time spectra; however, we note that the measurement is affected by contamination from H$\beta$ and line blending, which led to these two different velocity estimates. From these velocities, we inferred a lower limit on the $M_{ej} = 2.45 \mathrm{M_{\odot}}$ and an upper limit of $M_{ej}=7.36 \mathrm{M_{\odot}}$. In the fitting, we allowed the ejecta mass to vary beyond these bounds, since the adopted opacity $\kappa$, typical for fully ionised ejecta, provides only a lower limit on $M_{ej}$. The CSM mass $M_{CSM}$ was allowed to vary between $3$ and $20$ $\mathrm{M}_{\odot}$ (see Sect.~\ref{CSM_prop} for the estimation). 

The remaining free parameters include a noise parameter $\sigma$ (0.001 to 1), the gamma-ray opacity coefficient ($\kappa_{\gamma}$) of the SN ejecta (0.1 to $10^{4}~\mathrm{cm}^{2}~\mathrm{g}^{-1}$), the host galaxy hydrogen column density ($n_H$; $10^{16}$ to $10^{22} ~\mathrm{cm}^{-3}$), and the minimum temperature ($t_{min}$; 1000 to 50000 K). The full set of these 12 parameters, along with the inferred values, is provided in Table \ref{tab:mosfit}. The parameter ranges used can be found in Table~\ref{tab:param}.

The resulting fits are shown in Fig. \ref{fig:mosfit}, with the corresponding corner plots in Figs. \ref{fig:corner_0} and \ref{fig:corner_2}. Neither the $s=0$ (shell) nor $s=2$ (wind) CSM models adequately match the late-time decline of SN~2017dio. In both cases, we also see a very high $n_H$, which is in disagreement with no host galaxy extinction being visible in the spectra (a value of $10^{21.3}$ would already result in $A_V\approx 1$; \citealt{host_ex}). This suggests that the actual CSM density profile is more complex than a simple wind or shell configuration. 

We note that the model assumes homologous expansion and radiation-pressure dominance. It also assumes constant opacity, while in a more realistic scenario the opacity is depth- and time-dependent \citep{Cha_2012}.

The $s=0$ case, prefers lower ejecta mass and higher CSM mass, while $s=2$ has the opposite. The ejecta velocity is lower for $s=2$, but both cases are close to the lower velocity that we got from spectral fitting. The inner radius, R$_0$, is similar in the two cases, but density is lower in $s=0$. The $\mathrm{{^{56}Ni}}$ fraction is larger in the $s=0$ case, yielding a $\mathrm{{^{56}Ni}}$ mass $\mathrm{M({^{56}Ni})}$ of 0.14 $\mathrm{M_{\odot}}$, but due to the higher ejecta mass the $s=2$ model gives a slightly higher value of 0.2 $\mathrm{M_{\odot}}$.

\section{Infrared excess modelling}
\label{IR modelling}

The IR excess of SN~2017dio is clearly visible in the light curves and SEDs at late times. In this section we investigate the origin of the IR excess and the properties of the dust responsible for it. An IR excess in SNe is typically created by either pre-existing or newly formed dust. In the former scenario, the IR excess is created when the pre-existing dust is heated by the SN radiation and then re-emits in the IR bands (so-called IR echo; e.g. \citealt{Graham83}; \citealt{Graham86}; \citealt{Dwek1983}, \citeyear{Dwek1985}). The temporal evolution of the IR echo depends on the spatial distribution of dust. In the case of a spherically symmetric shell distribution, the IR radiation from the pre-existing dust appears suddenly following the SN and remains at roughly constant brightness for a duration of $t_{{\rm d}} = 2 r_{\rm{shell}}/c \sim 200$ days $(r_{\rm{shell}}/0.1 \rm{pc})$, where $r_{\rm{shell}}$ is the distance of the dust shell from the SN site of explosion and $c$ is the speed of light \citep[e.g.][]{Maeda+15}. 

The newly formed dust is typically considered to appear in either the inner parts of the SN ejecta (e.g. \citealt{dustejecta1}; \citealt{Nozawa}; \citealt{Dwek_2007}; \citealt{Sarangi}) or in the cool dense shell located between the forward and reverse shocks of the CSM interaction region (e.g. \citealt{Pozzo2004}; \citealt{Mattila2008}; \citealt{Gall_2011}; \citealt{Gall2018}). The estimated mass of newly formed dust at early times in the inner ejecta of Type II SNe is typically of the order of a few $\times 10^{-4}$ M$_{\odot}$ \citep{Shahbandeh2023}. Newly formed dust is sometimes accompanied by observational features such as asymmetries in emission-line profiles and/or accelerated light-curve declines at the shorter wavelengths (e.g. \citealt{Pozzo2004}; \citealt{Gall_2014}; \citealt{Bevan19}; \citealt{Shahbandeh2023}), depending on the spatial configuration of the emitting regions and the dust.

\subsection{IR echo from pre-existing dust}

\begin{figure}
\centering
\vspace{-0.3cm}
  \resizebox{\hsize}{!}{\includegraphics{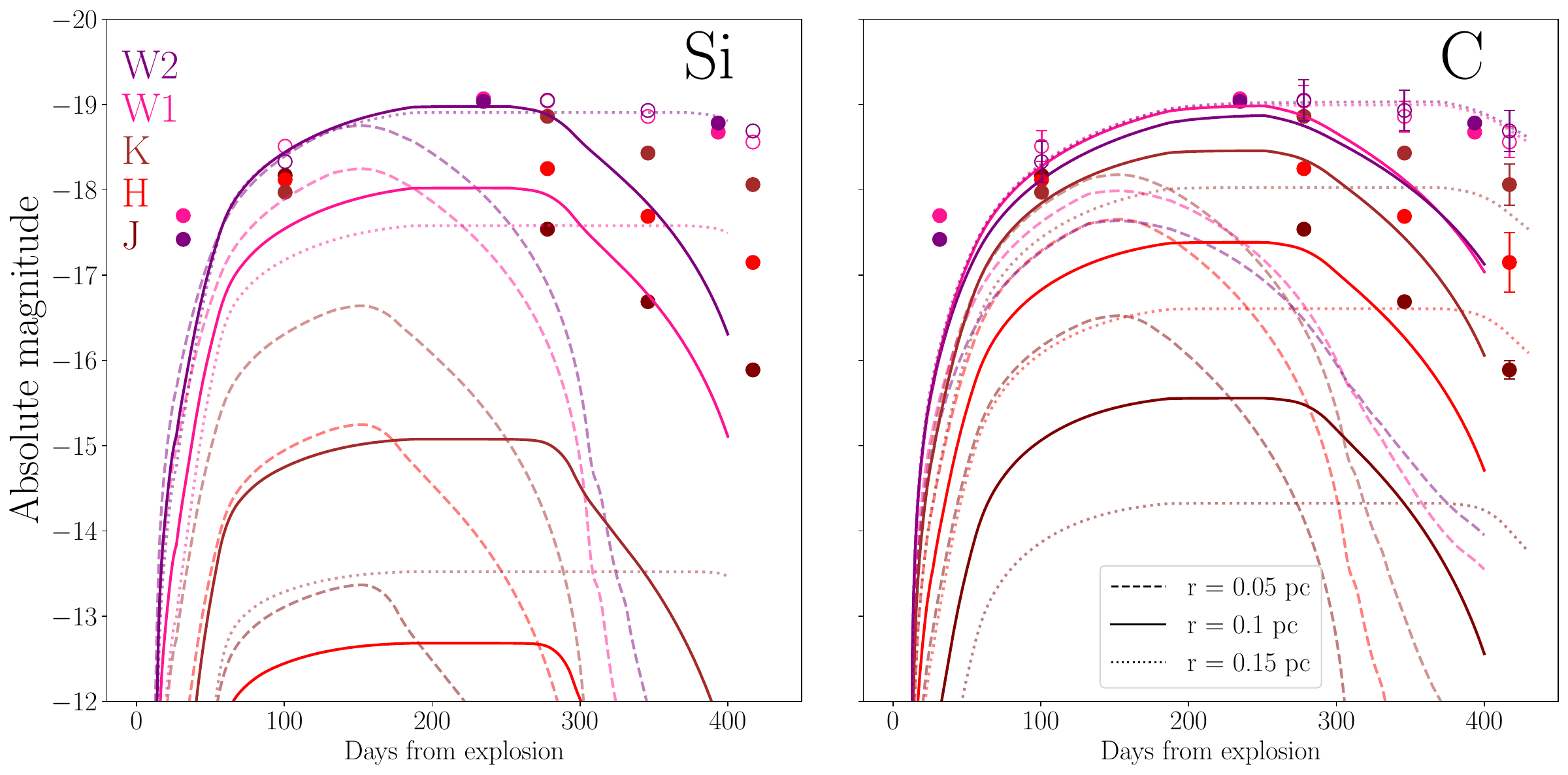}}
  \vspace{-0.2cm}
  \caption{Results of IR echo modelling using a thin shell at three different radii for both silicon and graphite dust. The radii used are $r = 0.05$~pc (dashed lines), $r=0.1$~pc (solid lines), and $r=0.15$~pc (dotted line). The IR photometry is plotted on top of the fits. Empty markers are interpolated data to match the NIR epochs.}
  \label{fig:nir}
\end{figure}

An IR excess, with a BB temperature of $\sim 1500$ K, was detected from $\sim 280$ to $\sim 760$ days after the explosion (see Sect. \ref{IR-excess}). To test whether this emission could arise from an IR echo, we modelled the IR excess assuming a spherically symmetric, thin dust shell, following the analytical method of \citet{Maeda+15}. The input SN radiation was taken from the bolometric luminosity and BB temperature derived from optical data, excluding the IR. We adopted the astronomical silicate and graphite dust models described in \cite{Nagao17}. Here, the dust grains are assumed to be spherical. Their size distribution follows the Mathis, Rumpl, and Nordsieck (MRN) power law, $\propto a^{-3.5}$, where $a$ is the radius of a dust grain \citep{Mathis1977}, with minimum and maximum grain radii of 0.05 and 0.2 $\mu$m, respectively.

We computed IR echoes for a range of shell radii ($r_{\rm shell}$) and dust masses ($M_{\rm dust}$) for both silicate and graphite compositions. The resulting fits are plotted in Fig. \ref{fig:nir}. While the observed duration of the IR excess is reasonably reproduced by models with $r_{\rm shell} \sim 0.1$ pc, the observed \textit{JHK} luminosities for this radius are much brighter than the predicted ones. The duration of the IR excess is roughly set by the light-travel time, $2 r_{\rm shell}/c$, implying $r_{\rm shell} \sim 0.13$ pc for a 300-day timescale. In contrast, since the observed IR excess temperature ($\sim 1500$ K) is close to the dust evaporation temperature, the dust responsible for the emission must reside near the expected evaporation radii. Here, the dust evaporation radii, estimated from radiative equilibrium \citep[Eq.~4 in][]{Maeda+15} with evaporation temperatures of $T_{\rm evp} = 1900$ K for graphite and 1500 K for silicate \citep{dust_evap}, of $\sim 0.017$ and $\sim 0.033$ pc, respectively, are much smaller. Therefore, an IR echo from pre-existing dust cannot simultaneously reproduce both the duration and temperature of the IR excess. Thus, we conclude that the majority of the IR excess is from newly formed dust.

From this IR echo analysis, we place an upper limit on the CSM mass at the dust evaporation radius, since most of the IR excess is expected to originate from newly formed dust. If the nearby CSM responsible for the early-time optical radiation extends to larger radii, dust in the CSM within the evaporation radius would have been destroyed by the SN radiation, while dust beyond this radius could produce an IR echo. The requirement that the echo from a dust shell at the dust evaporation radius should not be brighter than the observed IR excess in any band or at any epoch limits the dust mass at this radius to be $4 \times10^{-5}$ and $0.001$ M$_{\odot}$, corresponding to $2.4\times10^{-5}$ and $6\times10^{-4}$ M$_{\odot}$ yr$^{-1}$ of the gas mass-loss rates for carbon and silicate dust, respectively.

\subsection{Newly formed dust}

To estimate the newly formed dust masses, we used

\begin{equation}
    M_{dust} = \frac{L_{dust}}{4\sigma_{SB}\kappa_{abs}T^{4}_{dust}}
,\end{equation}where $L_{dust}$ is the bolometric luminosity of the dust, $\sigma_{SB}$ is the Stefan-Boltzmann constant, and $\kappa_{abs}$ is the absorption coefficient $\sim 10^3~ \mathrm{cm^2~g^{-1}}$ for graphite dust and $\sim 2\times10^2~ \mathrm{cm^2~g^{-1}}$ for silicate, assuming grain size $0.1~\mu$m and temperature $\sim 1000 ~$K \citep{Reynolds2025}. Based on the temperature and dust luminosity from BB fitting (see Sect.~4.2), we expect the observed IR excess to be dominated by the radiation from newly formed dust with an increasing mass from $\sim 0.001$ to $\sim 0.005$ M$_{\odot}$ in the case of graphite dust or $\sim 0.005$ to $\sim 0.02$ M$_{\odot}$ in the case of silicate dust, assuming no contribution from an IR echo. Such a large amount of dust has been observed in some interacting SNe, where the IR excess has been attributed to a combination of newly formed dust and an IR echo \citep[e.g.][]{Li22,Reynolds2025, Singh_26}. We note that the dust mass estimations for the last two epochs are affected by lack of data in NIR. The corresponding dust masses and temperatures for each epoch are listed in Table \ref{tab:dust_mass}.

The BB radius for the IR excess emission at $\sim 280$ days after the explosion is estimated to be $\sim 3\times 10^{16}$ cm (0.01 pc) based on the temperature and luminosity of the IR excess (see also Sect.~4). If the dust causing the IR excess formed inside the SN ejecta, the ejecta would need to expand to the observed BB radius in $\sim 280$ days, requiring a velocity of $\sim11,000 ~\mathrm{km~s^{-1}}$. Since this is unlikely, the newly formed dust was more likely produced in the cool dense shell in the CSM interaction regions rather than in the SN ejecta itself. This is also consistent with the fact that such a huge amount of dust formation is rarely seen in typical Type Ic SNe.

\section{Discussion}\label{discussions}

\subsection{SN ejecta properties}

In this section we compare the values for the SN ejecta inferred with MOSFiT with the ones we estimated with calculations and with values from the literature. Following Arnett's rule, we estimated the $\mathrm{{^{56}Ni}}$ mass to be $\sim 0.43\pm0.05~ \mathrm{M_{\odot}}$. This value is higher than the mean for non-interacting Type Ic (0.26 $\mathrm{M_{\odot}}$) and Type Ic-BL (0.31 $\mathrm{M_{\odot}}$) found in \cite{SESNE_ni}, who also used Arnett's rule in the estimation. So even if Arnett's rule tends to overestimate the $^{56}$Ni mass, SN~2017dio still produces more $^{56}$Ni than most non-interacting SESNe or the interaction affects the light curves during the bump and thus our estimation of the $^{56}$Ni mass. The modelling done by MOSFiT gives $\mathrm{{^{56}Ni}}$ mass $0.14~ \mathrm{M{_\odot}}~ (0.21~ \mathrm{M{_\odot}}$) for $s=0$ ($s=2$). Both values are consistent with the mean masses for SESNe. 

We estimated the range of $\mathrm{M_{ej}}$ to be $2.4-7.4~\mathrm{M_{\odot}}$ (see Sect.~\ref{LC_modelling}). The estimated lower limit ejecta mass is in good agreement with the mean ejecta masses for non-interacting Ic and Ic-BL SNe found in \cite{Taddia2018}. Our ejecta mass range for SN~2017dio is smaller than the value found by \cite{Shi_2024}, who modelled the light curves up to 100 days using a CSM+Ni model. The ejecta mass inferred with MOSFiT with $s=0$, 2.3 M${_\odot}$, is close to our estimated lower limit, while the ejecta mass, 5.4 M${_\odot}$, inferred with $s=2$ is 
slightly higher, but still fits our estimated range.

\subsection{CSM properties and mass-loss history}\label{CSM_prop}

The optical spectra and light curves of SN~2017dio suggest the presence of the nearby CSM ($\sim10^{15}$ cm; <0.001 pc) from the interaction. We estimated the mass loss for the CSM with 
\begin{equation}
    \mathrm{\dot{M}}= 2\frac{L_{H\alpha}}{\epsilon_{H\alpha}}\frac{v_{wind}}{v^3_{shock}},
\end{equation}
where L$_{H\alpha}$ is the luminosity of H$\alpha$, and $\epsilon_{H\alpha}$ is the efficiency factor 0.01 \citep{mass_loss_opt}. For $v_{wind}$, we assumed a velocity of $100~ \mathrm{km~s^{-1}}$; this is supported by no observations of excess in the red wing of the H$\alpha$, which would be expected if the $v_{wind}$ was over a few hundred $~\mathrm{km~s^{-1}}$ \citep{Huang}. For shock velocity, we assumed $v_{shock}=3500~ \mathrm{km~s^{-1}}$ based on results obtained for other SNe~IIn and suggested by the blue edge of the boxy component in the observed H$\alpha$ profile \citep[e.g.][]{Taddia2020}. We obtain a mass loss of $\sim 0.245~ \mathrm{M}_\odot~\mathrm{yr}^{-1}$ for the CSM. We estimate the mass-loss rate for all epochs of spectra and find that the mass-loss rate reaches $\sim 0.25~\mathrm{M}_\odot~\mathrm{yr}^{-1}$ at 90 days, when the interaction is strongest, but the typical value is $\sim~0.06~ \mathrm{M}_\odot~\mathrm{yr}^{-1}$. We estimated the CSM mass by integrating over the different mass-loss rates and find ${M}_{\mathrm{CSM}} \approx 8 ~\mathrm{M_{\odot}}$.

The delayed emergence of the CSM interaction component in the optical light curves (see Sect.~\ref{sec:photometric_properties}) indicates a relatively CSM-free cavity in the nearest regions of the SN progenitor, and thus a minimal mass-loss activity shortly before the explosion. The optical excess in the light curves appeared at $\sim 25$ days. With the assumption of $v_{ej} = 6000 ~\mathrm{km~s^{-1}}$ (see Sect.~\ref{LC_modelling}), the size of this low-CSM cavity corresponds to a distance of $\sim 1.3 \times10^{15}$ cm (87 AU; $4.2 \times10^{-4}$ pc). With a wind velocity of 100 $\mathrm{km~s^{-1}}$, the mass loss must have decreased significantly $\sim4$ years before the explosion. Due to the optical light curves and the spectra being powered by the interaction at late times, we know that there is CSM that has been formed even $\sim65$ years before the explosion, corresponding to a distance of $2.1 \times10^{16} $cm (1400 AU; 0.007 pc).

The echo modelling put an upper limit for the gas mass-loss rates around the dust evaporation radius to be $2.4\times10^{-5}$ and $6\times10^{-4}$ M$_{\odot}$ yr$^{-1}$ for the carbon and silicate dust, respectively. The dust evaporation radii for these dust models are $\sim 0.017$ and $\sim 0.033$ pc, respectively. With the assumption of $100~ \mathrm{km~s^{-1}}$ of the wind velocity, these radii correspond to the mass-loss history of $\sim 150$ and $\sim 300$ years before the explosion, respectively. Thus, the mass-loss rate of the progenitor should have been increased from these low values (several $\times \sim 10^{-5}$ M$_{\odot}$ yr$^{-1}$) to the high value estimated from the light-curve analysis (several $\times \sim 10^{-2}$ M$_{\odot}$ yr$^{-1}$) over the last $\lesssim 150-300$ years before the explosion.

In conclusion, the progenitor system of SN~2017dio underwent one significant mass-loss event between $\sim4$ and $65$ years before the explosion, possibly even having some lower mass-loss 150-300 years before the explosion.

\subsection{Progenitor system and mass-loss mechanism}

Here we discuss the progenitor system and the mass-loss mechanism based on the inferred properties of the SN ejecta and CSM. \citetalias{K18} interpreted that the CSM was not created by a typical steady-state stellar wind, but that the mass loss was most likely driven by eruptions or binary interaction. \cite{Shi_2024} also suggests that the CSM could have been created during mass-transfer in binaries or by eruptions of a luminous blue variable star or pulsational pair instability. 

The central finding regarding the H‑rich CSM is that it likely originates from the companion star rather than from the progenitor itself. To create an SN Ic, the progenitor star needs to first lose its H-envelope and, after it, its He-envelope. Therefore, if the progenitor produced the H-rich CSM, several solar masses of He-rich CSM should also be there. However, this expectation conflicts with two observational facts: SN~2017dio displays strong Balmer lines from early phases, and its light curve shows no excess that would arise from interaction with several solar masses of He‑rich CSM prior to the interaction with the H‑rich CSM. Thus, the H-rich CSM should come from a H-rich companion star and be ejected likely by binary interaction. This is consistent with the inferred aspherical distribution of the CSM. Similar origins for the H-rich material have also been suggested for the transitional SN~2017ens \citep{2017ens}, SN~2018ijp \citep{Tartaglia} and SN~2019yvr \citep{yvr}.

The classification as Type Ic implies that the progenitor of SN~2017dio did not accrete much of the H-rich gas during the mass transfer from the companion. This might be due to a low accretion efficiency or from super-Eddington accretion causing much of the transferred material to be ejected from the system rather than being retained by the progenitor.

\section{Conclusions}
\label{summary}

We have presented photometric and spectroscopic data of SN~2017dio from $\sim2$ to $\sim400$ days after the explosion in the optical bands and to $\sim800$ days in the IR bands. SN~2017dio represents an interesting case of a Type Ic SN interacting with H-rich CSM. Its light curves show a bump before rising to the main peak. This bump is powered by the SN itself, and the peak is powered by interaction. The rise to the main peak is also consistent with the timing when the interaction starts to dominate the spectra and hides the underlying SN type. The optical light curves decline at a rate faster than the expected $^{56}$Co tail, consistent with the light curves being powered by the interaction at late times. SN~2017dio also shows long-lasting IR excess, especially in the MIR bands.

We have shown that the progenitor system of SN~2017dio underwent one massive mass-loss event $\sim4-65$ years prior to explosion; this event created the CSM at a distance of $\sim1\times 10^{15}$cm (interaction visible in the optical bands). The IR excess can be explained by the mass~of newly formed dust increasing from $\sim 0.001-0.005~ \mathrm{M_{\odot}}$ (carbon) or $\sim 0.005-0.02 ~\mathrm{M_{\odot}}$ (silicate). The IR echo modelling sets an upper limit of $4\times10^{-5}~ \mathrm{M_{\odot}}$ pre-existing dust at the dust evaporation radius (0.017 pc for carbon), implying a progenitor mass-loss rate of $2.4 \times 10^{-5}~ \mathrm{M_{\odot}~yr^{-1}}$ roughly 170 years before the explosion. This indicates rapidly increasing mass loss ahead of the explosion.

Due to SN~2017dio being a He-poor SN, but still interacting with H-rich CSM, we argue that the CSM originates from a binary companion and was lost during mass transfer from the companion to the progenitor star. This could be due to super-Eddington accretion or otherwise a very low accretion efficiency. SN~2017dio provides us with a great opportunity to study the final years or months of massive stars in binaries.

\section*{Data availability}

All spectra are publicly available through the WISeREP archive: https://www.wiserep.org/object/2845

\begin{acknowledgements}
We thank the anonymous referee for the comments and suggestions that have helped us to improve the paper.
C.H. acknowledges financial support from Magnus Ehrnrooth Foundation.
C.P.G. acknowledges financial support from the Secretary of Universities and Research (Government of Catalonia) and by the Horizon 2020 Research and Innovation Programme of the European Union under the Marie Sk\l{}odowska-Curie and the Beatriu de Pin\'os 2021 BP 00168 programme, the support from the Spanish Ministerio de Ciencia e Innovaci\'on (MCIN) and the Agencia Estatal de Investigaci\'on (AEI) 10.13039/501100011033 under the PID2023-151307NB-I00 SNNEXT project, from Centro Superior de Investigaciones Cient\'ificas (CSIC) under the PIE project 20215AT016 and the program Unidad de Excelencia Mar\'ia de Maeztu CEX2020-001058-M, and from the Departament de Recerca i Universitats de la Generalitat de Catalunya through the 2021-SGR-01270 grant. 
H.K. and T.N. were funded by the Research Council of Finland projects 324504, 328898, 35301, and 340613.
S.M. acknowledges financial support from the Research Council of Finland project 350458.
T.K. acknowledges support from the Research Council of Finland project 360274.
K.M. acknowledges support from the Japan Society for the Promotion of
Science (JSPS) KAKENHI grant (JP24KK0070, JP24H01810). The work is partly
supported by the JSPS Open Partnership Bilateral Joint Research Projects
between Japan and Finland (K.M and H.K; JPJSBP120229923).
C.G. is supported by a Villum Experiment grant (VIL69896), from VILLUM FONDEN.
T.M.R is part of the Cosmic Dawn Center (DAWN), which is funded by the Danish National Research Foundation under grant DNRF140. T.M.R acknowledges support from the Research Council of Finland project 350458.
A.M.G. acknowledges financial support from grant PID2023-152609OA-
I00, funded by the Spanish Ministerio de Ciencia, Innovación y Universidades
(MICIU), the Agencia Estatal de Investigación (AEI, 10.13039/501100011033),
and the European Union’s European Regional Development Fund (ERDF).
M.D. Stritzinger is funded by the Independent Research Fund Denmark (IRFD, grant number  10.46540/2032-00022B).

This research has made use of the NASA/IPAC Infrared Science Archive, which is funded by the National Aeronautics and Space Administration and operated by the California Institute of Technology.

This work has made use of data from the Asteroid Terrestrial-impact Last Alert System (ATLAS) project. The Asteroid Terrestrial-impact Last Alert System (ATLAS) project is primarily funded to search for near earth asteroids through NASA grants NN12AR55G, 80NSSC18K0284, and 80NSSC18K1575; byproducts of the NEO search include images and catalogs from the survey area. This work was partially funded by Kepler/K2 grant J1944/80NSSC19K0112 and HST GO-15889, and STFC grants ST/T000198/1 and ST/S006109/1. The ATLAS science products have been made possible through the contributions of the University of Hawaii Institute for Astronomy, the Queen’s University Belfast, the Space Telescope Science Institute, the South African Astronomical Observatory, and The Millennium Institute of Astrophysics (MAS), Chile.

The CSS survey is funded by the National Aeronautics and Space
Administration under Grant No. NNG05GF22G issued through the Science
Mission Directorate Near-Earth Objects Observations Program.  The CRTS
survey is supported by the U.S.~National Science Foundation under
grants AST-0909182 and AST-1313422.

The Pan-STARRS1 Surveys (PS1) and the PS1 public science archive have been made possible through contributions by the Institute for Astronomy, the University of Hawaii, the Pan-STARRS Project Office, the Max-Planck Society and its participating institutes, the Max Planck Institute for Astronomy, Heidelberg and the Max Planck Institute for Extraterrestrial Physics, Garching, The Johns Hopkins University, Durham University, the University of Edinburgh, the Queen's University Belfast, the Harvard-Smithsonian Center for Astrophysics, the Las Cumbres Observatory Global Telescope Network Incorporated, the National Central University of Taiwan, the Space Telescope Science Institute, the National Aeronautics and Space Administration under Grant No. NNX08AR22G issued through the Planetary Science Division of the NASA Science Mission Directorate, the National Science Foundation Grant No. AST-1238877, the University of Maryland, Eotvos Lorand University (ELTE), the Los Alamos National Laboratory, and the Gordon and Betty Moore Foundation.

The Liverpool Telescope is operated on the island of La Palma by Liverpool John Moores University in the Spanish Observatorio del Roque de los Muchachos of the Instituto de Astrofisica de Canarias with financial support from the UK Science and Technology Facilities Council.

Based on observations made with the Nordic Optical Telescope, owned in collaboration by the University of Turku and Aarhus University, and operated jointly by Aarhus University, the University of Turku and the University of Oslo, representing Denmark, Finland and Norway, the University of Iceland and Stockholm University at the Observatorio del Roque de los Muchachos, La Palma, Spain, of the Instituto de Astrofisica de Canarias. The NOT data were obtained as part of the NUTS2 collaboration.

\end{acknowledgements}

%
\bibliographystyle{aa} 
\bibliography{Files/lahteet} 
%

\clearpage
\begin{appendix}  
\section{Figures}
\FloatBarrier

\begin{figure}[h]
  \resizebox{\hsize}{!}{\includegraphics{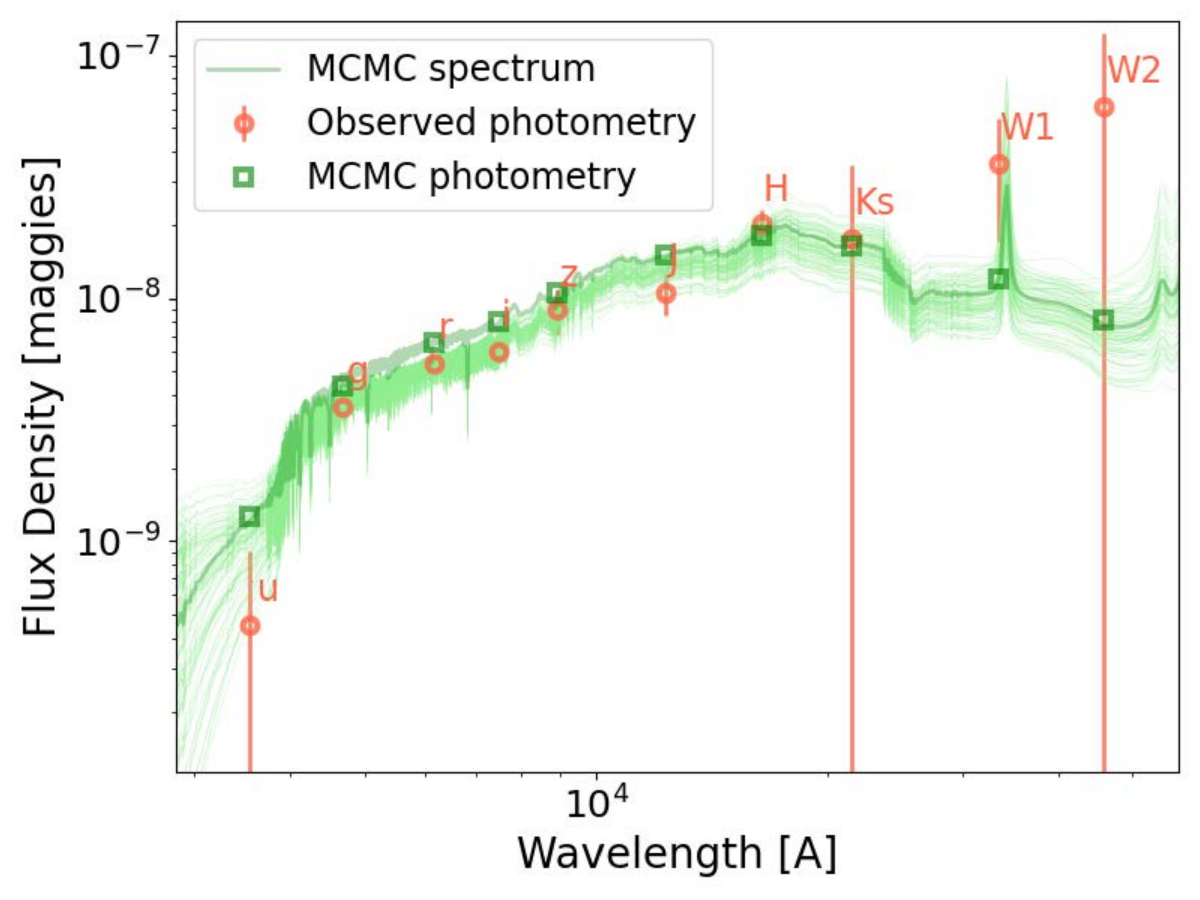}}
  \caption{SED of the host galaxy derived with \textsc{prospector}. The observed host galaxy photometry are marked with orange circles and the Markov chain Monte Carlo photometry with green squares.}
  \label{fig:SED_gal}
\end{figure}

\begin{figure}[h]
  \resizebox{\hsize}{!}{\includegraphics{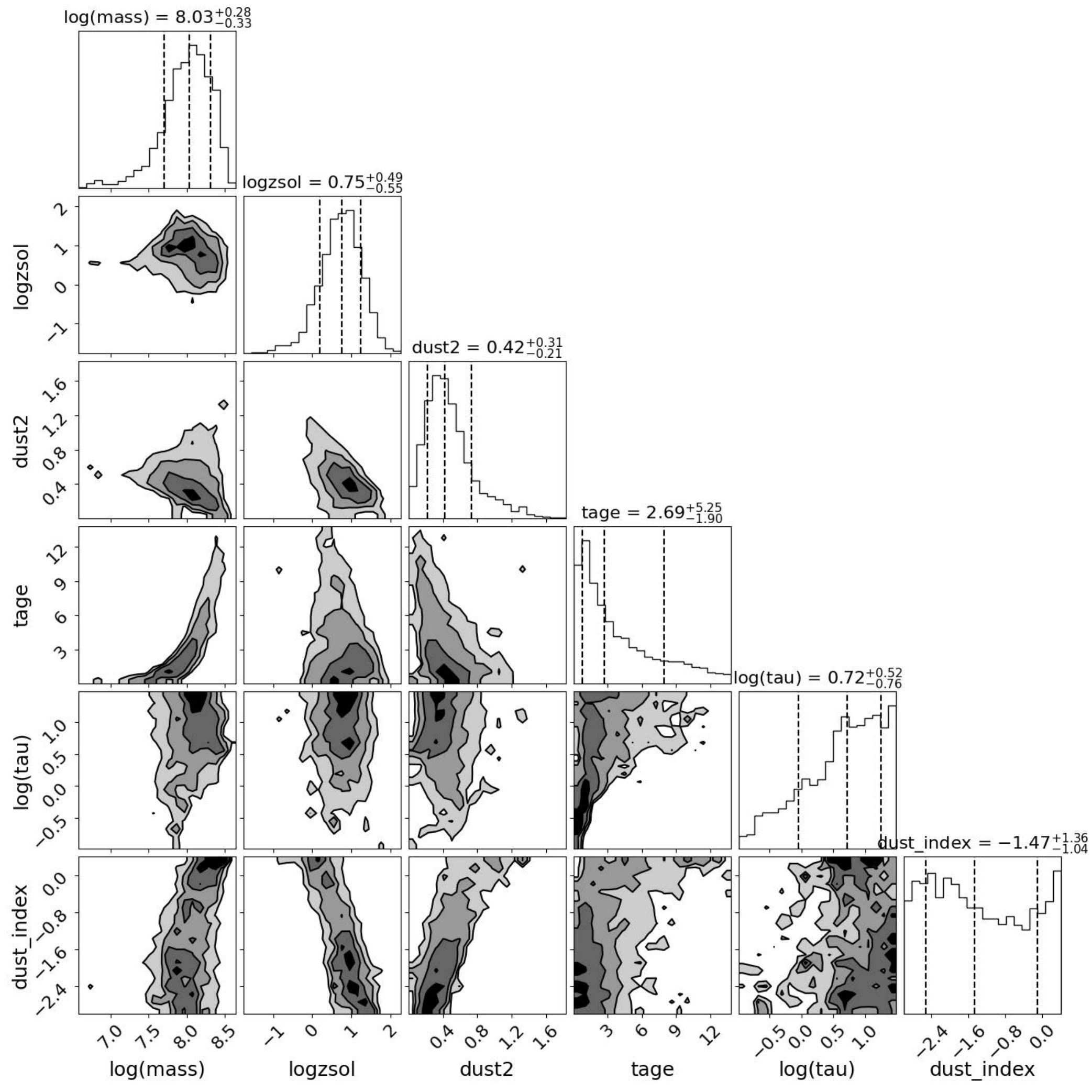}}
  \caption{Corner plot for the host galaxy parameters.}
  \label{fig:mcmc_gal}
\end{figure}

\begin{figure}
  \resizebox{\hsize}{!}{\includegraphics{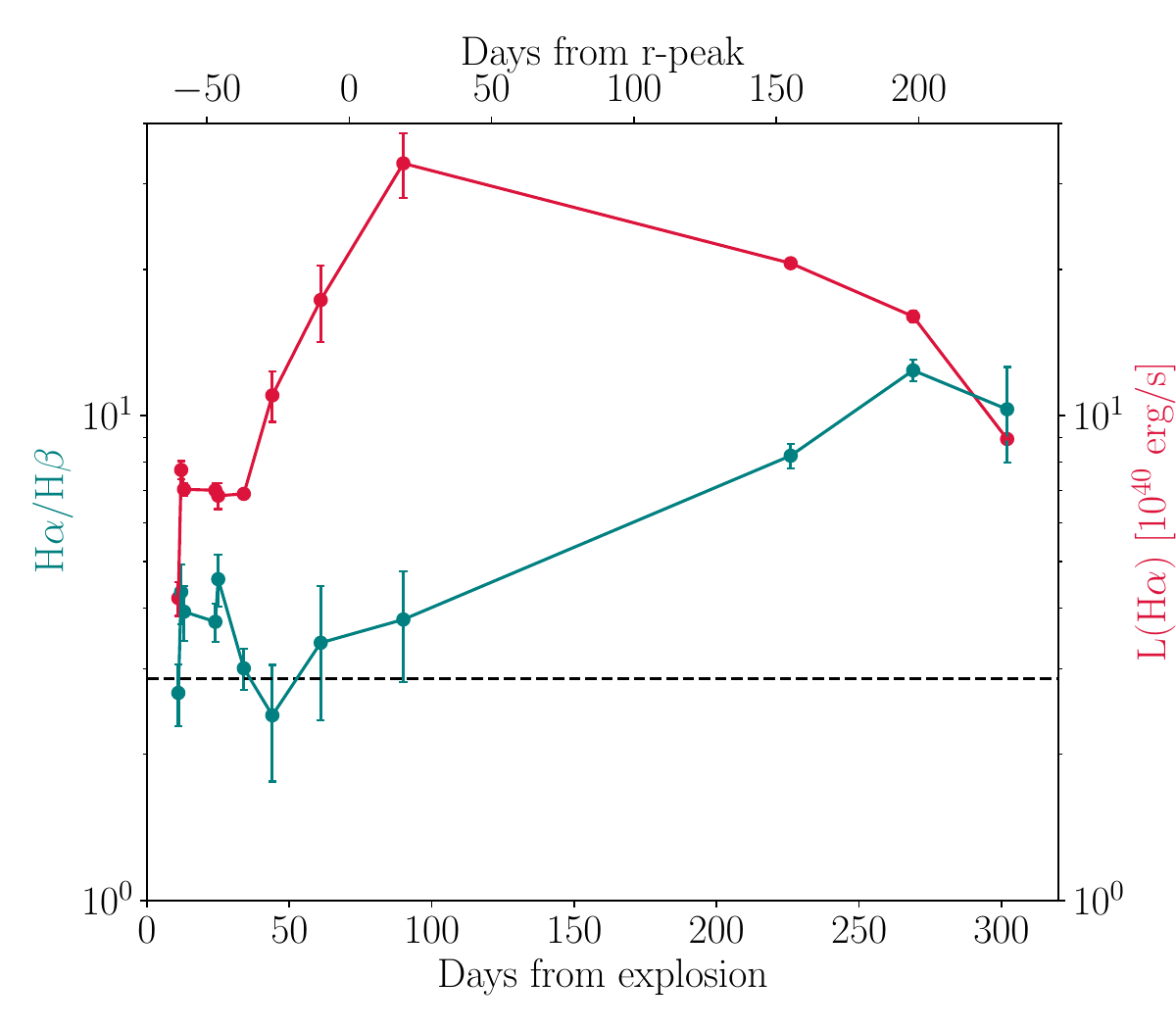}}
  \caption{Ratio of H$\alpha$ and H$\beta$ (in green) and the H$\alpha$ luminosity (in red). The units of luminosity are in $10^{40}$ erg/s. The y-axis is in log-scale, and the x-axis is in days from the explosion (\textit{bottom}) and days from the $r$ peak (\textit{top}). The dashed grey line shows the expected Balmer decrement value for Case B.}
  \label{fig:HaHb_ratio}
\end{figure}

\begin{figure}
  \resizebox{\hsize}{!}{\includegraphics{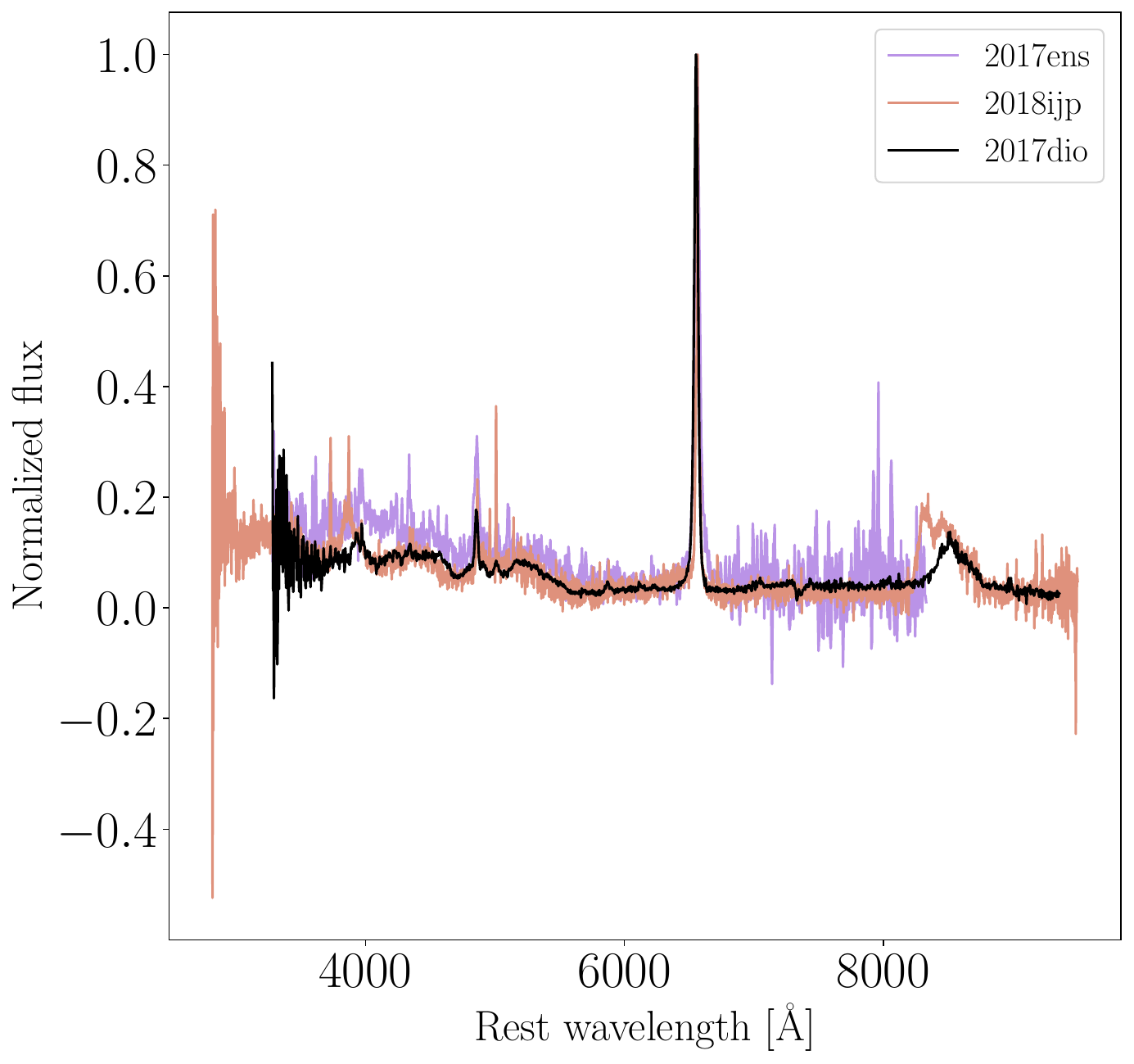}}
  \caption{Spectral comparison of SN~2017dio (black), SN~2017ens (violet), and SN~2018ijp (brown) at approximately 300-400 days after the explosion.}
  \label{fig:close_comp}
\end{figure}

\begin{figure*}
\centering
\resizebox{0.62\hsize}{!}{\includegraphics{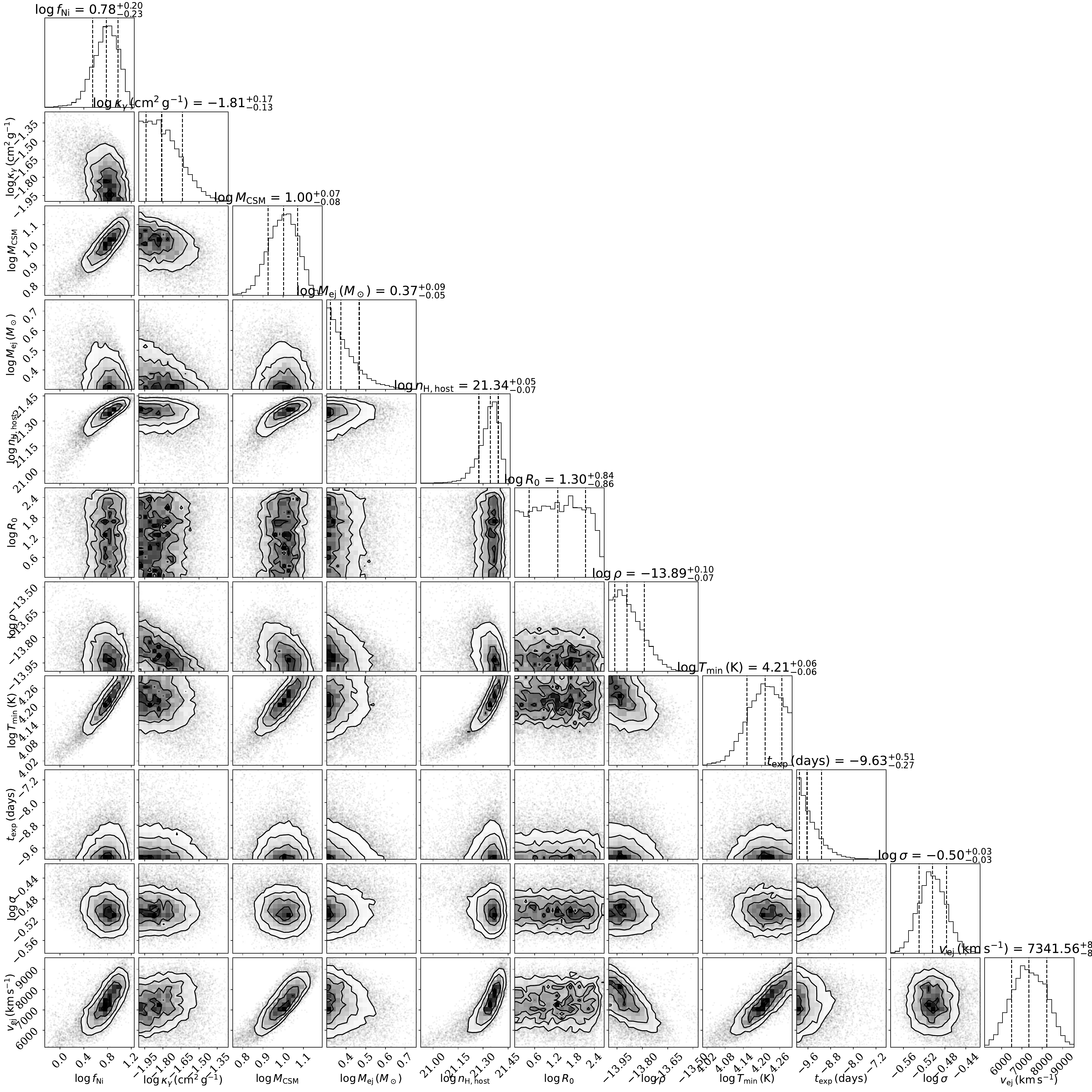}}
\caption{Corner plot for s=0.}
\label{fig:corner_0}
\end{figure*}

\begin{figure*}
\centering
\resizebox{0.62\hsize}{!}{\includegraphics{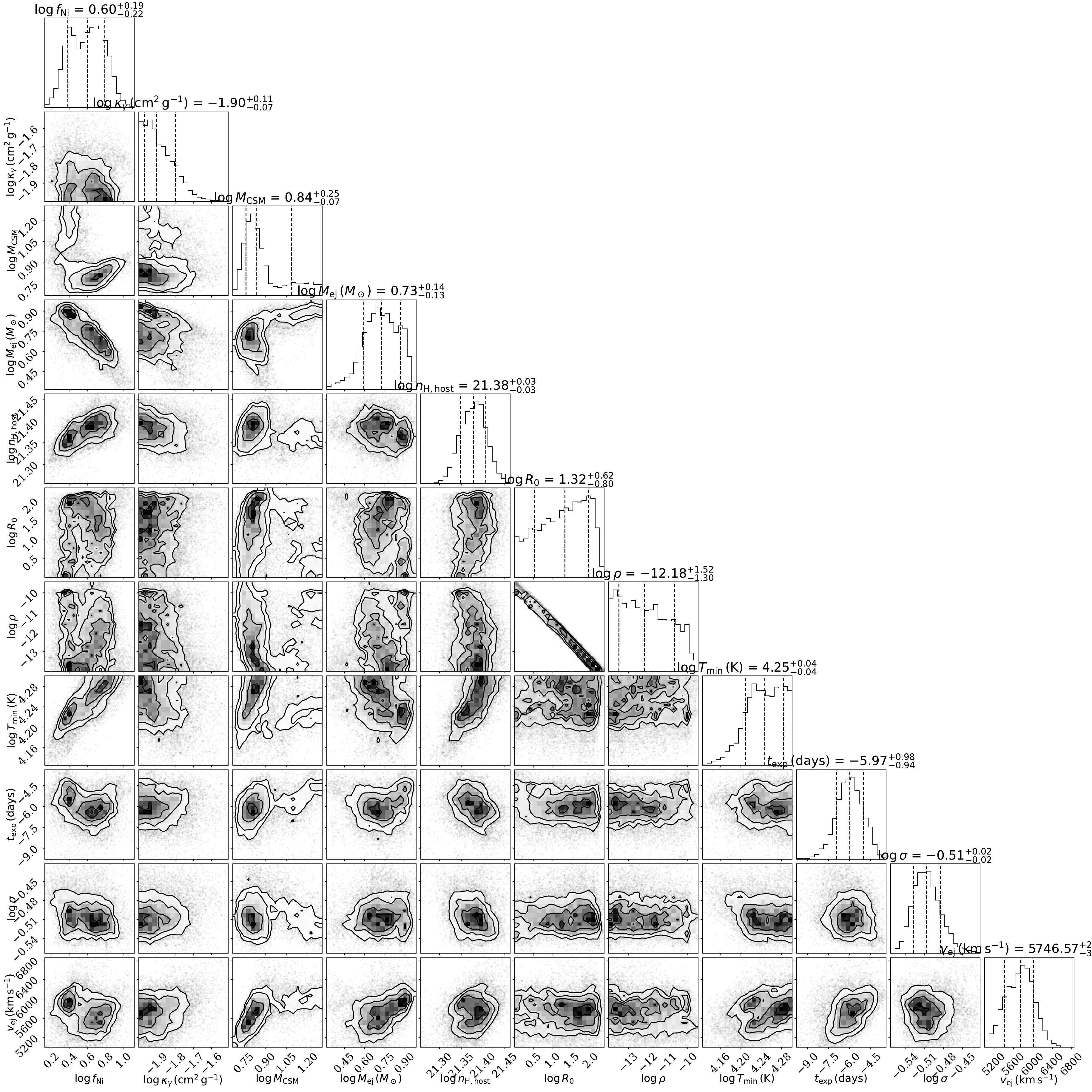}}
\caption{Corner plot for s=2.}
\label{fig:corner_2}
\end{figure*}

\FloatBarrier

\section{Tables}

\begin{table}[h]
\small
    \caption{Photometric data taken with ATLAS in orange (o) and cyan (c) bands, both in the AB system.}
    \label{tab:atlas}
    \begin{tabular}{lrcc}
\hline\hline 
 Date (UT)  &   MJD  &                 o &                 c \\
\hline
 2017-04-20 &  57863.51 &  19.05 $\pm$ 0.29 &    --          \\
 2017-04-25 &  57868.50 &  18.48 $\pm$ 0.17 &     --          \\
 2017-04-26 &  57869.50 &   --             &  18.37 $\pm$ 0.05 \\
 2017-05-07 &  57880.34 &   18.05 $\pm$ 0.1 &    --          \\
 2017-05-08 &  57881.35 &  17.98 $\pm$ 0.05 &    --          \\
 2017-05-09 &  57882.36 &  17.94 $\pm$ 0.09 &    --          \\
 2017-05-12 &  57885.34 &  17.63 $\pm$ 0.06 &    --          \\
 2017-05-14 &  57887.31 &  17.82 $\pm$ 0.04 &    --          \\
 2017-05-16 &  57889.36 &  17.92 $\pm$ 0.04 &    --          \\
 2017-05-18 &  57891.30 &  17.86 $\pm$ 0.04 &    --          \\
 2017-05-20 &  57893.36 &     --            &  17.97 $\pm$ 0.03 \\
 2017-05-23 &  57896.26 &  17.69 $\pm$ 0.04 &     --         \\
 2017-05-26 &  57899.25 &   17.6 $\pm$ 0.02 &     --         \\
 2017-05-27 &  57900.25 &  17.58 $\pm$ 0.03 &     --         \\
 2017-05-28 &  57901.36 &     --            &  17.62 $\pm$ 0.02 \\
 2017-05-30 &  57903.26 &   17.5 $\pm$ 0.03 &     --         \\
 2017-06-03 &  57907.26 &  17.45 $\pm$ 0.05 &     --         \\
 2017-06-05 &  57909.26 &  17.48 $\pm$ 0.05 &     --         \\
 2017-06-09 &  57913.27 &  17.37 $\pm$ 0.07 &     --         \\
 2017-06-20 &  57924.27 &  17.26 $\pm$ 0.02 &     --         \\
 2017-06-24 &  57928.27 &  17.25 $\pm$ 0.02 &     --         \\
 2017-06-28 &  57932.27 &  17.26 $\pm$ 0.02 &     --         \\
 2017-07-02 &  57936.29 &  17.29 $\pm$ 0.02 &     --         \\
 2017-07-06 &  57940.29 &  17.33 $\pm$ 0.03 &     --         \\
 2017-07-08 &  57942.29 &  17.28 $\pm$ 0.04 &     --         \\
 2017-07-12 &  57946.30 &  17.31 $\pm$ 0.03 &     --         \\
 2017-07-14 &  57948.30 &  17.34 $\pm$ 0.04 &     --         \\
 2017-10-29 &  58055.64 &  18.52 $\pm$ 0.08 &     --         \\
 2017-11-02 &  58059.64 &  18.53 $\pm$ 0.1  &     --         \\
 2017-11-06 &  58063.64 &  18.57 $\pm$ 0.2  &     --         \\
 2017-11-08 &  58065.62 &  18.64 $\pm$ 0.16 &     --         \\
 2017-12-14 &  58101.65 &  19.63 $\pm$ 0.13 &     --         \\
 2017-12-18 &  58105.65 &  19.8 $\pm$ 0.21  &     --         \\
\hline
\end{tabular}
\tablefoot{The data are stacked by date.}
\end{table}

\begin{table}[h]
\small
\caption{Photometric data from PAN-STARRS in the $w_{P1}$ band.} 
\label{tab:Pan-STARRS} 
\begin{tabular}{ l l}
\hline\hline 
MJD & $w_{P1}$  \\
\hline 
57868.39 &18.43 $\pm$ 0.03\\
57896.32 &17.79 $\pm$ 0.02\\
58118.57 &20.08 $\pm$ 0.13\\
58132.51 &20.10 $\pm$ 0.04\\
58159.44 &20.91 $\pm$ 0.14\\
58184.37 &21.18 $\pm$ 0.12\\
58187.37 &21.18 $\pm$ 0.12\\
\hline 
\end{tabular}
\tablefoot{The data have been stacked by date.}
\end{table}

\begin{table*}[h]
\tiny
\center
\caption{Photometric data in the optical bands.}
\label{tab:optical_1}
\begin{tabular}{llrccccccc}
    \hline\hline 
        Date (UT) & Telescope & MJD  & B & g & V & r & i & z \\ 
    \hline
        2017-04-22 & CSS & 57865 & -- & 18.95 $\pm$ 0.17 & 19.13 $\pm$ 0.13 & -- & -- & -- \\ 
        2017-04-28 & CSS & 57871 & -- & 18.43 $\pm$ 0.13 & 18.91 $\pm$ 0.18 & -- & -- & -- \\ 
        2017-05-01 & CSS & 57874 & -- & 18.19 $\pm$ 0.09 & 18.16 $\pm$ 0.09 & -- & -- & -- \\ 
        2017-05-02 & LT+IO:O & 57875.89 & -- & 18.25 $\pm$ 0.11 & -- & 17.97 $\pm$ 0.09 & 18.13 $\pm$ 0.04 & 18.13 $\pm$ 0.11 \\ 
        2017-05-13 & LT+IO:O & 57885.99 & -- & 18.32 $\pm$ 0.29 & -- & 17.98 $\pm$ 0.43 & 17.69 $\pm$ 0.67 & -- \\ 
        2017-07-02 & CSS & 57936  & -- & -- & 17.10 $\pm$ 0.00 & -- & -- & -- \\ 
        2017-10-22 & NOT+ALFOSC& 58048.22 & 18.39 $\pm$ 0.01 & -- & -- & -- & -- & -- \\
        2017-10-22 & NOT+ALFOSC& 58048.23  & -- & -- & 18.07 $\pm$ 0.07 & 17.94 $\pm$ 0.01 & 18.48 $\pm$ 0.0 & 17.25 $\pm$ 0.19 \\
        2017-10-27 & LT+IO:O & 58053.25 & 18.52 $\pm$ 0.22 & -- & -- & -- & -- & -- \\
        2017-10-27 & LT+IO:O & 58053.26  & 18.52 $\pm$ 0.22 & 18.28 $\pm$ 0.15 & 18.03 $\pm$ 0.13 & 18.12 $\pm$ 0.21 & 18.35 $\pm$ 0.19 & 17.71 $\pm$ 0.23 \\ 
        2017-11-01 & LT+IO:O & 58058.25  & 18.55 $\pm$ 0.05 & 18.21 $\pm$ 0.04 & 18.18 $\pm$ 0.05 & 18.1 $\pm$ 0.07 & 18.47 $\pm$ 0.08 & 17.71 $\pm$ 0.07 \\ 
        2017-11-12 & LT+IO:O & 58069.24  & 18.80 $\pm$ 0.09 & 18.46 $\pm$ 0.13 & 18.03 $\pm$ 0.12 & 18.51 $\pm$ 0.29 & 18.82 $\pm$ 0.24 & 17.58 $\pm$ 0.6 \\ 
        2017-11-15 & LT+IO:O & 58072.21  & 18.75 $\pm$ 0.10 & 18.55 $\pm$ 0.08 & 18.49 $\pm$ 0.08 & 18.37 $\pm$ 0.08 & 18.75 $\pm$ 0.08 & 18.0 $\pm$ 0.1 \\ 
        2017-11-18 & LT+IO:O & 58075.20  & 18.88 $\pm$ 0.06 & 18.57 $\pm$ 0.0 & 18.53 $\pm$ 0.03 & 18.47 $\pm$ 0.03 & 18.81 $\pm$ 0.02 & 18.04 $\pm$ 0.03 \\
        2017-11-21 & LT+IO:O & 58078.21  & 19.14 $\pm$ 0.49 & -- & -- &-- & -- & -- \\ 
        2017-11-21 & LT+IO:O & 58078.22  & -- & 18.63 $\pm$ 0.17 & 18.71 $\pm$ 0.33 & 18.51 $\pm$ 0.15 & 18.59 $\pm$ 0.25 & 18.04 $\pm$ 0.47 \\
        2017-11-24 & LT+IO:O & 58081.19  & 19.01 $\pm$ 0.21 & -- & 18.61 $\pm$ 0.11 & -- & -- & -- \\ 
        2017-11-24 & LT+IO:O & 58081.20  & -- & 18.69 $\pm$ 0.07 & -- & 18.51 $\pm$ 0.07 & 19.03 $\pm$ 0.15 & 18.19 $\pm$ 0.16 \\ 
        2017-12-15 & LT+IO:O & 58102.27  & 19.49 $\pm$ 0.10 & -- & -- & -- & -- & -- \\
        2017-12-15 & LT+IO:O & 58102.28  & -- & 19.21 $\pm$ 0.08 & 19.19 $\pm$ 0.09 & 19.04 $\pm$ 0.07 & 19.29 $\pm$ 0.12 & 18.64 $\pm$ 0.12 \\
        2017-12-18 & LT+IO:O & 58105.28& 19.73 $\pm$ 0.18 & -- & 19.17 $\pm$ 0.12 & -- & -- & -- \\
        2017-12-18 & LT+IO:O & 58105.29& --  & 19.24 $\pm$ 0.09 & -- & 19.08 $\pm$ 0.11 & 19.49 $\pm$ 0.18 & 18.53 $\pm$ 0.12 \\ 
        2017-12-20 & LT+IO:O & 58107.25  & 19.58 $\pm$ 0.07 & 19.34 $\pm$ 0.05 & 19.32 $\pm$ 0.05 & 19.11 $\pm$ 0.05 & -- & -- \\
        2017-12-20 & LT+IO:O & 58107.26  & -- & -- & -- & -- & 19.5 $\pm$ 0.12 & 18.7 $\pm$ 0.08 \\ 
        2017-12-21 & LT+IO:O & 58108.24 & 19.67 $\pm$ 0.11 & 19.38 $\pm$ 0.06 & 19.36 $\pm$ 0.08 & 19.17 $\pm$ 0.07 & 19.48 $\pm$ 0.1 & 18.8 $\pm$ 0.11 \\ 
        2017-12-25 & LT+IO:O & 58112.23 & -- & 19.45 $\pm$ 0.04 & -- & 19.26 $\pm$ 0.06 & -- & -- \\
        2017-12-25 & LT+IO:O & 58112.24 &-- & -- & -- & -- & 19.56 $\pm$ 0.1 & 18.85 $\pm$ 0.11 \\
        2017-12-26 & LT+IO:O & 58113.22&--  & 19.49 $\pm$ 0.04 & -- & 19.29 $\pm$ 0.05 & 19.64 $\pm$ 0.14 & -- \\ 
        2017-12-26 & LT+IO:O & 58113.23 &--  & -- & -- & -- & -- & 18.87 $\pm$ 0.09 \\ 
        2017-12-28 & LT+IO:O & 58115.12 & 19.73 $\pm$ 0.14 & 19.58 $\pm$ 0.02 & 19.60 $\pm$ 0.13 & 19.35 $\pm$ 0.02 & 19.66 $\pm$ 0.01 & -- \\
        2017-12-28 & LT+IO:O & 58115.13&--  & -- & -- & -- & -- & 18.96 $\pm$ 0.01 \\
        2017-12-28 & LT+IO:O & 58115.21& --  & 19.58 $\pm$ 0.02 & -- & -- & -- & -- \\ 
        2017-12-28 & LT+IO:O & 58115.22 &--  & -- & -- & 19.35 $\pm$ 0.02 & 19.66 $\pm$ 0.01 & 18.96 $\pm$ 0.01 \\ 
        2017-12-29 & LT+IO:O & 58116.26&-- & 19.56 $\pm$ 0.08 & -- & 19.32 $\pm$ 0.05 & -- & -- \\
        2017-12-29 & LT+IO:O & 58116.27&--   & -- & -- & -- & 19.64 $\pm$ 0.1 & 18.97 $\pm$ 0.09 \\
        2018-01-01 & LT+IO:O & 58119.09&--  & 19.42 $\pm$ 0.2 & -- & 19.25 $\pm$ 0.14 & 19.63 $\pm$ 0.15 & -- \\
        2018-01-01 & LT+IO:O & 58119.10&--  & -- & -- & -- & -- & 19.0 $\pm$ 0.15 \\
        2018-01-11 & LT+IO:O & 58129.20 & 20.24 $\pm$ 0.12 & -- & 19.95 $\pm$ 0.06 & 19.69 $\pm$ 0.07 & -- & 19.28 $\pm$ 0.12 \\
        2018-01-11 & LT+IO:O & 58129.21 & -- & 19.94 $\pm$ 0.06 & -- & -- & 20.00 $\pm$ 0.10 & -- \\
        2018-01-20 & LT+IO:O & 58138.24 & 20.43 $\pm$ 0.11 & -- & 20.12 $\pm$ 0.08& 19.67 $\pm$ 0.09 & -- & 19.41 $\pm$ 0.15 \\
        2018-01-20 & LT+IO:O & 58138.25 & -- & 20.13 $\pm$ 0.07 & -- & -- & 19.93 $\pm$ 0.11 & -- \\
        2018-01-22 & LT+IO:O & 58140.13  & -- & 19.94 $\pm$ 0.06 & -- & 19.64 $\pm$ 0.09 & 20.02 $\pm$ 0.15 & -- \\
        2018-01-22 & LT+IO:O & 58140.13 & -- & -- & -- & -- & -- & 19.35 $\pm$ 0.2 \\ 
        2018-01-28 & LT+IO:O & 58146.16  & 20.61 $\pm$ 0.34 & -- & 20.12 $\pm$ 0.24 & -- & -- & 19.6 $\pm$ 0.2 \\
        2018-01-28 & LT+IO:O & 58146.17  & -- & 20.15 $\pm$ 0.21 & -- & 19.9 $\pm$ 0.18 & 20.19 $\pm$ 0.29 & -- \\
        2018-02-14 & NOT+ALFOSC& 58163.26 & -- & 20.71 $\pm$ 0.0 & -- & 19.69 $\pm$ 0.52 & -- & -- \\ 
        2018-04-11 & NOT+ALFOSC& 58219.11  & -- & 21.36 $\pm$ 0.04 & -- & 20.19 $\pm$ 0.58 & -- & -- \\
        2018-05-19 & NOT+ALFOSC & 58257.00  & -- & 21.41 $\pm$ 0.02 & -- & 20.36 $\pm$ 0.55 & 20.84 $\pm$ 0.02 & -- \\ 
        2018-05-19 & NOT+ALFOSC & 58257.01 & -- & -- & -- & -- & -- & 20.83 $\pm$ 0.02 \\ 
\hline
\end{tabular}
\tablefoot{\textit{griz} bands are in the AB system and \textit{VB} bands in the Vega system.}
\end{table*}

\begin{table*}[h]
\small
\caption{Photometric IR data in the Vega system.}
\label{tab:NIR_phot}
\centering
\begin{tabular}{llrccccc}

\hline\hline 
 Date (UT) & Telescope & MJD & J&H & K & W1 & W2\\
\hline
   &WISE&57893.64&--&--&--&15.63$\pm$ 0.18 &15.33 $\pm$ 0.3\\
    &WISE&58096.75 &--&--&--&14.26 $\pm$ 0.09 & 13.71 $\pm$ 0.14\\
    2018-01-22&NOT+NOTCAM&58141.15 &17.58 $\pm$ 0.01 &-- & --&--&--\\
    2018-01-22&NOT+NOTCAM&58141.17 &--&16.38 $\pm$ 0.01& -- &--&--\\
    2018-01-22&NOT+NOTCAM&58141.19 &--&--& 15.34 $\pm$ 0.01&--&--\\
    2018-03-31&NOT+NOTCAM&58208.88 &--&--& 15.77 $\pm$ 0.01&--&--\\
    2018-03-31&NOT+NOTCAM&58208.89 &--&16.94 $\pm$ 0.01& --&--&--\\
    2018-03-31&NOT+NOTCAM&58208.91 &18.43 $\pm$ 0.01 &--& --&--&--\\
    &WISE& 58254.13 &--&--&--&--&13.97 $\pm$ 0.15\\
    &WISE&58257.20&--&--&--&14.65 $\pm$ 0.13&--\\
    2018-06-10&NOT+NOTCAM&58279.89 &--& & 16.14 $\pm$ 0.24&--&--\\
    2018-06-10&NOT+NOTCAM&58279.91 &--& 17.48 $\pm$ 0.35& -- &--&--\\
    2018-06-10&NOT+NOTCAM&58279.93 &19.23 $\pm$ 0.11 & -- & -- &--&--\\
    &WISE&58462.44 &--&--&--&15.77 $\pm$ 0.16 & 14.84 $\pm$ 0.25\\
    &WISE&58621.13 &--&--&--&16.52 $\pm$ 0.32 & 15.20 $\pm$ 0.34\\
\hline
\end{tabular}
\tablefoot{WISE data are stacked and show the mean MJD.}
\end{table*}

\begin{table}[h]
\caption{Log of the spectroscopic observations of SN~2017dio.} 
\label{tab:spec}
\centering
\begin{tabular}{c l l l l l l} 
\hline\hline 
Date (UT) & Telescope+Instrument&Phase (d) & MJD & Grism & $\Delta\lambda$ (Å) & Range (Å)\\
\hline 
2017-04-30 & ESO-NTT+EFOSC2-NTT&11 & 57873  & 13& 21.2 & 3685-9315\\
2017-05-01 & ESO-NTT+EFOSC2-NTT& 12& 57874  & 11& 15.8& 3380-7520\\
2017-05-01 & ESO-NTT+EFOSC2-NTT& 12& 57874 & 16& 16& 6015-10320\\
2017-05-02 & NOT+ALFOSC&13 & 57875 & 4 & 16.2 & 3200-9600\\
2017-05-13 & LT+SPRAT&24 & 57886 & &18 & 4000-8000\\
2017-05-14 & NOT+ALFOSC&25 & 57887 & 4 & 16.2 & 3200-9600\\
2017-05-23 & NOT+ALFOSC&34 & 57896 & 4 & 16.2 & 3200-9600\\
2017-06-02 & ESO-NTT+EFOSC2-NTT&44 & 57906 & 11&15.8&3380-7520 \\
2017-06-02 & ESO-NTT+EFOSC2-NTT&44 & 57906 & 16& 16& 6015-10320\\
2017-06-19 & NOT+ALFOSC&61 & 57923 & 4 & 16.2 & 3200-9600\\
2017-07-18 & NOT+ALFOSC&90 & 57952 & 4 & 16.2 & 3200-9600\\
2017-12-01 & NOT+ALFOSC&226 & 58088 & 4 & 16.2 & 3200-9600\\
2018-01-13 & NOT+ALFOSC&269 & 58131 & 4 & 16.2 & 3200-9600\\
2018-02-15 & NOT+ALFOSC&302 & 58164 & 4 & 16.2 & 3200-9600\\
\hline 
\end{tabular}
\tablefoot{Phase is given as days after the explosion.}
\end{table}

\begin{table}[h]
\small
\caption{Parameter ranges used in MOSFiT modelling.} 
\label{tab:param} 
\centering 
\begin{tabular}{l l l}
\hline\hline 
Parameter & Range & Unit \\
\hline 
 $f_{Ni}$ & 0.01 -- 20 &\% \\
$M_{CSM}$ & 3 -- 20 & $\mathrm{M}_{\odot}$ \\
$M_{ej}$  & 2 -- 25 & $\mathrm{M}_{\odot}$  \\
$R_0$ & 1 -- 500 & AU  \\
$\rho_0$ & $10^{-15}$ -- $10^{-10}$ & $\mathrm{g}~\mathrm{cm}^{-3}$ \\
$v_{ej}$ & 5000 -- 18000 & $\mathrm{km}~ \mathrm{s}^{-1}$ \\
$t_{exp} $& -10 -- 0 & days  \\
$\sigma $& 0.001 -- 1 & mag  \\
$\kappa_{\gamma} $& 0.01 -- 0.06 & $\mathrm{cm}^{2}~\mathrm{g}^{-1}$  \\
$n_H$ & $10^{16}$ -- $10^{22}$ & cm$^{-2}$   \\
$T_{min} $& 1000 -- 50000 & K \\
\hline 
\end{tabular}
\end{table}

\begin{table}[h]
\small
\caption{BB luminosity, temperature, and estimated dust masses for SN~2017dio at different epochs.} 
\label{tab:dust_mass} 
\centering 
\begin{tabular}{l l l l l}
\hline\hline 
Phase (d) & $L_{BB}~ (10^{42}$ erg/s) & $T_{BB}$ (K) & $M_{Si}$ ($10^{-3}~M_{\odot}$) & $M_C$ ($10^{-3}~M_{\odot}$)\\
\hline 
278 & 2.53 $\pm$ 1.6 & 1500 $\pm$ 150& 5.6 $\pm$ 3& 1.1 $\pm$ 0.5 \\
346 & 1.93 $\pm$ 1.7 &1470 $\pm$ 25 & 4.6 $\pm$ 4& 0.9 $\pm$ 0.8\\
417 & 1.43 $\pm$ 1.38 &1340 $\pm$ 10 & 5 $\pm$ 0.2& 1 $\pm$ 0.04 \\
600 & 0.56 $\pm$ 0.3&790 $\pm$ 90 & 16 $\pm$ 5& 3 $\pm$ 1 \\
759 & 0.4 $\pm$ 0.2&650 $\pm$100 & 24 $\pm$ 2& 5 $\pm$ 0.5\\
\hline 
\end{tabular}
\tablefoot{Phase is given as days after the explosion.}
\end{table}

\end{appendix}

\end{document}